# Temperature-dependent conformations of exciton-coupled Cy3 dimers in double-stranded DNA

Loni Kringle,[1] Nicolas P. D. Sawaya,[2] Julia Widom,[3] Carson Adams,[1] Michael G. Raymer,[4] Alán Aspuru-Guzik,[2,5] and Andrew H. Marcus[1,*]

[1.] Department of Chemistry and Biochemistry, Center for Optical, Molecular and Quantum Science, University of Oregon, Eugene, OR 97403, USA

[2.] Department of Chemistry and Chemical Biology, Harvard University, Cambridge, MA 02138, USA

[3.] Department of Chemistry, University of Michigan, Ann Arbor, Michigan 48109, United States

[4.] Department of Physics, Center for Optical, Molecular and Quantum Science, University of Oregon, Eugene, OR 97403, USA

[5.] Senior Fellow, Canadian Institute for Advanced Research (CIFAR), Toronto, Ontario M5G 1Z8, Canada

[*]e-mail: ahmarcus@uoregon.edu

## Abstract

Understanding the properties of electronically interacting molecular chromophores, which involve internally coupled electronic-vibrational motions, is important to the spectroscopy of many biologically relevant systems. Here we apply linear absorption, circular dichroism (CD), and two-dimensional fluorescence spectroscopy (2DFS) to study the polarized collective excitations of excitonically coupled cyanine dimers $(Cy3)_2$ that are rigidly positioned within the opposing sugar-phosphate backbones of the double-stranded region of a double-stranded (ss) – single-stranded (ss) DNA fork construct. We show that the exciton-coupling strength of the $(Cy3)_2$-DNA construct can be systematically varied with temperature below the ds – ss DNA denaturation transition. We interpret spectroscopic measurements in terms of the Holstein vibronic dimer model, from which we obtain information about the local conformation of the (Cy3)$_2$ dimer, as well as the degree of static disorder experienced by the Cy3 monomer and the (Cy3)$_2$ dimer probe locally within their respective DNA duplex environments. The properties of the (Cy3)$_2$-DNA construct we determine suggest that it may be employed as a useful model system to test fundamental concepts of protein-DNA interactions, and the role of electronic-vibrational coherence in electronic energy migration within exciton-coupled bio-molecular arrays.

## I. Introduction

A long-standing problem in molecular spectroscopy is to understand the roles of nuclear vibrations in the electronic structure of interacting molecules (1-13). Since the early work of Förster, Kasha, Fulton and Gouterman (2, 14-16), it has been recognized that the absorption spectra of interacting molecules can appear strikingly different than that of the constituent monomers, particularly when the monomer spectrum exhibits a pronounced vibronic progression, which is due to the coupling between electronic and vibrational motion (17-19). Such situations are important to the spectroscopic properties of molecular aggregates including biological and artificial light harvesting arrays (7, 13, 20-22).

Time-resolved ultrafast experiments that probe the excited-state dynamics of photosynthetic antenna complexes suggest that quantum coherence might contribute to the energy transfer mechanism of these systems (20-24). Recent studies show that spectroscopic signatures of quantum coherence can be understood by considering the role of vibrations; specifically, the presence of spatially delocalized electronic-vibrational states (10-13, 22, 25). These and other experiments have stimulated new ideas for molecular design principles that utilize resonant intermolecular electronic (exciton) coupling in combination with intra-molecular electronic-vibrational (vibronic) coherences as a resource to achieve enhancements in energy transfer efficiency (26, 27). In order to test these principles experimentally, it is useful to develop molecular systems for which the exciton coupling strength can be varied while intramolecular parameters such as coupling between electronic and vibrational modes are maintained constant.

In the following work, we study the effects of varying exciton coupling strength on the vibronic transitions of a molecular dimer composed of two Cy3 chromophores incorporated into the sugar-phosphate backbone of the double-stranded (ds) region of a DNA replication fork construct (see Fig. 1). Such fluorescently labeled DNA constructs may be used to study detailed mechanisms of protein-DNA interactions. Cy3 is a commonly used fluorescent probe for biophysical studies of protein-DNA interactions (28-35), which may be placed at site-specific positions within single-stranded (ss) DNA using phosphoramidite chemical insertion methods (28, 32). By annealing two complementary DNA strands, each labeled internally with a single Cy3 chromophore, a DNA duplex can be formed with the resulting (Cy3)$_2$ dimer adopting a chiral conformation with approximately $D_2$ symmetry. The stability of the (Cy3)$_2$ dimer depends on the strength of complementary hydrogen bonds between opposing nucleic acid base pairs, which can be adjusted by varying temperature and solvent conditions (36). Previous absorption and circular dichroism (CD) studies of internally labeled Cy3 (and other cyanine dyes) in dsDNA indicate that the local movements of the chromophores are restricted in the lowest energy, all trans, electronic ground state (33, 35). Furthermore, in typical aqueous solutions the spectroscopic lineshapes of these systems appear to be inhomogeneously broadened, which implies the existence of a wide range of local structural environments experienced by the Cy3 probes at any instant. The concept of DNA 'breathing' (i.e., local structural fluctuations) (36) allows for local disordered regions of the sugar-phosphate backbone to persist on time scales much longer than the Cy3 excited state lifetime (< 1 ns) (28). Nevertheless, little is known about the conformations of such sub-states or the time scales of their inter-conversion.

We show that the temperature-dependent linear absorption and CD of the (Cy3)$_2$ dimer in dsDNA can be well characterized using the Holstein model (37), which we apply to a model with two conformational parameters: the inter-chromophore separation $R_{AB}$ and the twist angle $\phi_{AB}$. The Holstein Hamiltonian describes each molecular site as a two-level electronic system coupled to a single harmonic mode (5, 15, 16, 37, 38). The presence of a resonant electronic interaction between the two molecular sites (characterized by the parameter $J$) leads to the formation of a manifold of excited states composed of delocalized symmetric and anti-symmetric superpositions of electronic-vibrational tensor product states. Such 'effective state models' have been well established to describe the electronic properties of molecular dimers in the strong and intermediate exciton-coupling regimes (5, 7, 8, 11, 12, 15, 16, 38). The combination of absorption and CD spectroscopy is particularly well suited to resolve these states since absorptive transitions to symmetric and anti-symmetric excited states are orthogonally polarized, and thus contribute to the CD spectrum with opposite sign.

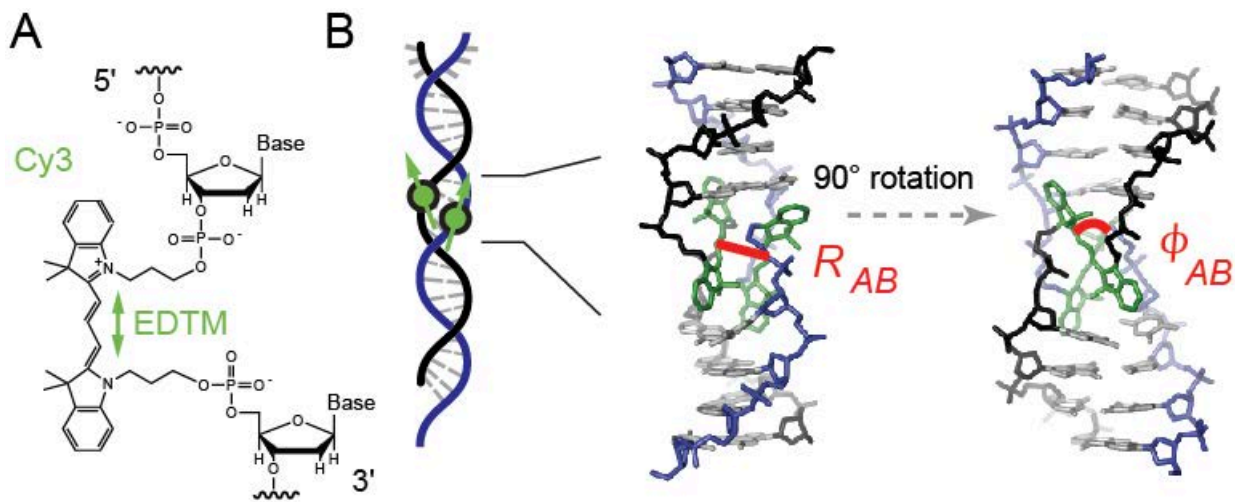

**Figure 1.** Model structure of the internally labeled (Cy3)$_2$ dimer in dsDNA. (***A***) The structural formula of the internally labeled Cy3 chromophore is shown with its insertion linkages to the 3' and 5' segments of the sugar-phosphate backbone of ssDNA. A green double-headed arrow indicates the orientation of the electric dipole transition moment (EDTM), which lies parallel to the plane of the trimethine bridge. (***B***) A dsDNA segment formed from two complementary DNA strands, which each contains an internally labeled Cy3 chromophore, serves as a scaffold to hold the (Cy3)$_2$ dimer in place. Space-filling structural models performed using the Spartan program (Wavefunction, Inc.) suggest that the dimer exhibits the same approximate $D_2$ symmetry as right-handed (B-form) helical dsDNA. The sugar-phosphate backbones of the conjugate strands are shown in black and blue, the bases in gray, and the Cy3 chromophores in green. Additional space-filling renderings of the structure are presented in Fig. S5. The structural parameters that define the local conformation of the (Cy3)$_2$ dimer are the inter-chromophore separation vector ***R***$_{AB}$ and the twist angle $\phi_{AB}$.

Förster classified the interaction strength of an exciton-coupled molecular aggregate according to the degree of distortion exhibited by its absorption and photoluminescence spectra in comparison to those of the constituent monomers. He identified three different coupling regimes: strong, intermediate, and weak (2). The strong coupling regime corresponds to a major redistribution of intensity between the various vibronic sub-bands of the aggregate absorption spectrum. In the intermediate coupling regime, the intensities of the vibronic bands of the aggregate are similar to those of the monomer, yet each vibronic feature is broadened due to these interactions. In the weak-coupling regime (often referred to as the Förster regime), the absorption spectrum of the coupled system is indistinguishable from that of a collection of uncoupled monomers. However, in a weakly coupled system, a local excitation may stochastically hop from one site to another (2). In this work, we show that by adjusting the temperature over the range 15 – 60 °C, which spans the pre-melting regime of the (Cy3)$_2$–dsDNA system, we may vary the *intermolecular* coupling strength $J$ over the range 530 – 450 cm$^{-1}$ while maintaining *intramolecular* parameters approximately constant, such as the monomer transition energy $\varepsilon_{eg}$, the vibrational frequency $\omega_0$, and the electron-vibrational coupling strength characterized by the Huang-Rhys parameter $\lambda^2$. In the following, we show that the structural parameters and degree of static disorder that characterizes the Hamiltonian of the system undergo systematic and physically meaningful changes as a function of temperature, which acts mainly to vary the inter-chromophore separation and twist angle through its destabilizing effects on local secondary structure of the DNA duplex. Because only the inter-chromophore properties of the (Cy3)$_2$-dsDNA construct are sensitive to temperature, the system may be employed as a useful experimental model to test fundamental concepts of protein-DNA interactions, and the role of electronic-vibrational coherence in electronic energy migration within exciton-coupled bio-molecular arrays.

## II. Experimental Methods

***Sample Preparation.*** The sequences and nomenclature of the internally labeled Cy3 ssDNA constructs used in this work are shown in Table I. Oligonucleotide samples were purchased from Integrated DNA Technologies (IDT, Coralville, IA) and used as received. We prepared solutions using a standard aqueous buffer of 10 mM Tris, 100 mM NaCl, and 6 mM MgCl$_2$, with concentration of 400 nM for our absorption and 2DFS measurements, and 1 μM for our CD measurements. We combined complementary oligonucleotide strands to form the Cy3 monomer and (Cy3)$_2$ dimer labeled DNA fork constructs, which contain both ds and ss regions. For both the monomer and dimer labeled samples, the probes were positioned deep in the duplex DNA region. The monomer labeled construct contained a thymine base (T) in the complementary strand position directly opposite to the Cy3 probe chromophore. Prior to the experiments, the sample solutions were annealed by heating to 95 °C for 3 minutes before they were allowed to slowly cool overnight.

**Table I.** Base sequences and nomenclature for the Cy3 monomer and (Cy3)$_2$ dimer DNA constructs used in these studies. The horizontal line indicates the regions of complementary base pairing.

| dsDNA construct | Nucleotide base sequence |
| --- | --- |
| Cy3 monomer | 3'-GTC AGT ATT ATA CGC TCy3C GCT AAT ATA TAC GTT TTT TTT TTT TTT TTT TTT TTT TTT TTT T-5'<br>5'-CAG TCA TAA TAT GCG A T  G CGA TTA TAT ATG CTT TTA CCA CTT TCA CTC ACG TGC TTA C-3' |
| (Cy3)$_2$ dimer | 3'-GTC AGT ATT ATA CGC TCy3C GCT AAT ATA TAC GTT TTT TTT TTT TTT TTT TTT TTT TTT TTT T-5'<br>5'-CAG TCA TAA TAT GCGACy3G CGA TTA TAT ATG CTT TTA CCA CTT TCA CTC ACG TGC TTA C-3' |

***Absorption and CD Measurements.*** Linear absorption measurements were carried out for each sample using a Cary 3E UV-Vis spectrophotometer, and CD measurements were performed using a Jasco model J-720 CD spectrophotometer. Both instruments were equipped with a computer-controlled temperature stage, which held the solutions in a 1 cm quartz cuvette. Absorption and CD spectra were measured over the range 200 – 700 nm to simultaneously examine the spectral region of the native bases (~275 nm) in addition to that of the Cy3 probe(s) (~540 nm). Room temperature (25 °C) absorption and CD spectra for the monomer and dimer labeled DNA constructs are shown over the visible spectral range in Fig. 2. Spectra corresponding to the ultraviolet absorbance and CD of the nucleobases confirmed that the ds regions of the DNA constructs adopted the anticipated Watson-Crick right-handed B-form conformation (see Fig. S1 of the SI) (39). The absorption spectrum of the Cy3 monomer DNA construct exhibits a progression of vibronic features with the first (0–0) peak centered at 549 nm (18,280 cm$^{-1}$). The vibronic progression is still present in the spectrum of the (Cy3)$_2$ dimer DNA construct. However, individual vibronic features of the dimer are broadened relative to those of the monomer, and the ratio of the 0–0 to 1–0 vibronic peak intensities has decreased relative to that of the monomer [$I_{mon}^{(0-0)}/I_{mon}^{(1-0)}$ = 1.60]. While the monomer CD signal is very weak (as expected), the dimer CD exhibits a progression of bisignate lineshapes (i.e. a change of sign within a given vibronic band), which is a signature of vibronic excitons in a chiral aggregate (3, 8, 40).

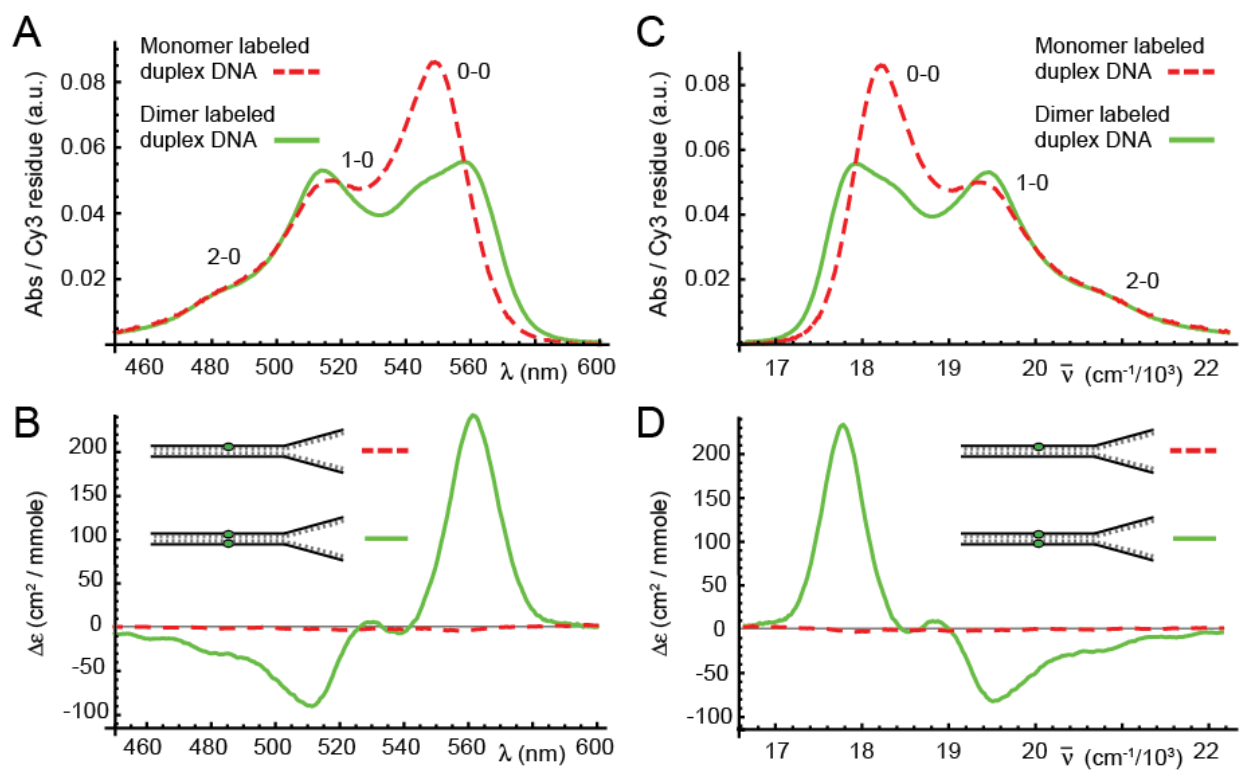

**Figure 2.** Room temperature (25°C) absorption (***A** & **C***) and CD (***B** & **D***) spectra for Cy3 monomer (dashed red) and dimer (solid green) labeled DNA constructs. Here $\Delta\varepsilon$ is the differential absorption of left and right circular polarized light. Nucleotide sequences and placement of the chromophore probes are indicated in Table 1. The spectra are shown as a function of optical wavelength (***A** & **B***) and as a function of wavenumber (***C** & **D***). The vibronic features of the monomer absorption spectra are labeled $n_e - 0$, where $n_e$ (= 0, 1, 2) indicates the vibrational occupancy of the electronically excited monomer.

***Two-Dimensional Fluorescence Spectroscopy (2DFS).*** To obtain an estimate of the homogeneous line widths for the two samples described in Table I, we performed phase-modulated 2DFS experiments at room temperature. These measurements were carried out as previously described (41-44). The four laser pulses were generated from a single high-repetition-rate non-collinear optical parametric amplifier (NOPA) with the excitation centered at 535 nm and a bandwidth of 16 nm for experiments performed on the $(Cy3)_2$ dimer DNA construct, and an excitation centered at 530 nm with 17 nm bandwidth for experiments performed on the Cy3 monomer DNA construct. Fluorescence was detected using a 570 – 616 nm band-pass filter (Semrock FF01-593/40-25), which served to reject scattered excitation light. To eliminate optical saturation effects, solutions were continuously circulated through the cuvette using a peristaltic pump. Pulses were compressed using a double-pass glass SF10 prism pair to compensate for dispersive media in the optical path preceding the sample, as previously described (41). Pulse widths were characterized by placing a beta-barium borate (BBO) frequency doubling crystal at the sample position, where a phase-modulated train of pulse-pairs was incident. The frequency-doubled signal output was detected using a lock-in amplifier, which was referenced to the ac carrier signal used to modulate the relative phase of the pulses. We thus minimized the pulse width $\Delta \tau_L$ by performing a pulse-pulse autocorrelation. We measured the laser bandwidth $\Delta \lambda_L =$ ~16 nm centered at $\lambda_L$ = 535 nm using an Ocean Optics mini-spectrometer. The measured time-bandwidth product was thus ~ $\Delta \tau_L (\Delta \lambda_L c/\lambda_L^2)$ ~ 0.53, which is within 20% of the optimal value (0.44) for Fourier-transform-limited Gaussian pulses. Results from these measurements are presented below.

## III. Theoretical Modeling

We implemented the well-known Holstein Hamiltonian, which has been previously applied to model the vibronic character of an electronically interacting cyanine dimer (11, 12). Because each Cy3 chromophore is rigidly attached at two insertion site positions within the DNA single strands, the conformational space available to the (Cy3)$_2$ dimer of the fully annealed DNA duplex is restricted. Simple van der Waals models suggested that the (Cy3)$_2$ dimer adopts a chiral conformation with approximately $D_2$ symmetry (see Fig. 1 and Fig. S5 of the SI). We refer to the monomer sites as $A$ and $B$, and specify the conformation by the inter-chromophore separation $R_{AB}$ and twist angle $\phi_{AB}$. In the following discussion, we refer to the coordinate system shown in Fig. 3.

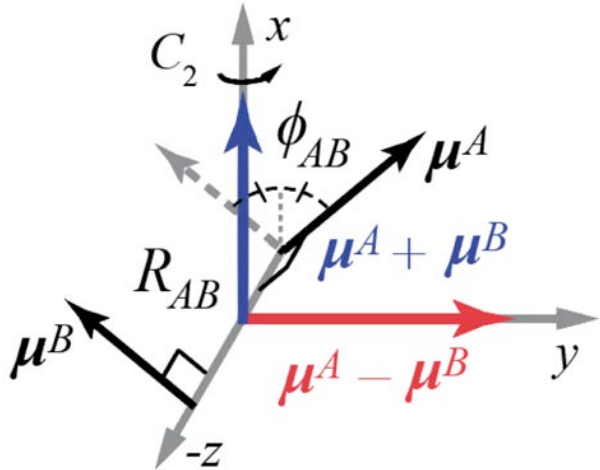

**Figure 3.** Cartesian coordinate system for the $AB$ dimer. Monomer EDTMs, $\boldsymbol{\mu}^A$ and $\boldsymbol{\mu}^B$, are separated by the distance $R_{AB}$, and twist angle $\phi_{AB}$. The monomer EDTMs are assumed to be perpendicular with respect to the $z$-axis. The symmetric exciton $\boldsymbol{\mu}^A + \boldsymbol{\mu}^B$ (shown in blue) and the anti-symmetric exciton $\boldsymbol{\mu}^A - \boldsymbol{\mu}^B$ (shown in red) are each oriented parallel to the $x$- and $y$-axes, respectively. The $x$-axis is an axis of $C_2$ symmetry, as indicated. (The $y$- and $z$-axes are similarly $C_2$ symmetry elements.) Note that the point dipole-dipole coupling strength $J \propto cos\phi_{AB}$ [see Eq. (6) below] undergoes a sign inversion at $\phi_{AB} = 90°$.

***Monomer Hamiltonian and Absorption Spectrum.*** We consider the EDTM (electric dipole transition moment) of each Cy3 chromophore to be aligned parallel to the long axis of its trimethine chain (see Fig. 1A) (28, 32). The monomer EDTM is defined as the matrix element $\boldsymbol{\mu}_{eg}^M = \langle e|\hat{\boldsymbol{\mu}}^M|g\rangle_M$, where the operator $\hat{\boldsymbol{\mu}}^M = |g\rangle_M \mu_{ge}\langle e| + |e\rangle_M \mu_{eg}\langle g|$ couples the ground electronic state $|g\rangle_M$ to the excited electronic state $|e\rangle_M$ with transition energy $\varepsilon_M = \varepsilon_{eg}$. Each monomer $M\ (=A,B)$ has its two-level electronic transition coupled to a single harmonic mode with frequency $\omega_0$ and generalized coordinate $q_M$. The identity operator for a monomer is given by the tensor product of electronic and vibrational state contributions: $\hat{I}_M = \hat{I}_M^{elec} \otimes \hat{I}_M^{vib}$, where $\hat{I}_M^{elec} = |g\rangle_M\langle g| + |e\rangle_M\langle e|$, $\hat{I}_M^{vib} = \sum_{n_g} |n_g\rangle_M\langle n_g|$, and where $n_g$ is the population number of vibrational excitations (phonons) in the ground electronic state. The identity operator for the composite $AB$ system is thus $\hat{I}_A \otimes \hat{I}_B$. In our numerical calculations described below we obtained convergent results using a maximum of six vibrational excitations per monomer, consistent with the findings of others (5, 12).

In the composite space of the $AB$ dimer, the Hamiltonian for each monomer is given by (10)

$$\hat{H}_M = \left\{ \frac{1}{2}[\hat{p}_M^2/m + m\omega_0^2 \hat{q}_M^2]|g\rangle_M\langle g| \right. \tag{1}$$
$$\left. + \left[\varepsilon_{eg} + \frac{1}{2}(\hat{p}_M^2/m + m\omega_0^2(\hat{q}_M - d)^2)\right]|e\rangle_M\langle e| \right\} \otimes \hat{I}_{M'\neq M}$$

where $\hat{q}_M$ and $\hat{p}_M$ are coordinate and momentum operators, respectively, for the monomer's internal vibration and $m$ is its reduced mass. Here we have taken the energy of the electronic ground state to be zero, and $d$ is the Franck-Condon displacement projected onto the vibrational coordinate $\hat{q}_M$ (see Fig. 4A).

The intensities of the absorptive transitions are determined by the square matrix elements $\left|\langle e|\langle n_e|\hat{\boldsymbol{\mu}}^M|g\rangle|n_g = 0\rangle_{A(B)}\right|^2 = |\boldsymbol{\mu}_{eg}^M|^2 |\langle n_e|0\rangle|^2$, where we have taken the initial state of the molecule to be both electronically and vibrationally unexcited. We use the Condon approximation, which assumes that the EDTM is unaffected by the vibrational mode. It follows

that the monomer absorption spectrum $\sigma^M_{H-abs}(\varepsilon)$ is the sum of homogeneous lineshapes associated with the individual vibronic transitions (45), given by

$$\sigma^M_{H-abs}(\varepsilon) = |\boldsymbol{\mu}^M_{eg}|^2 \sum_{n_e}^{\infty} |\langle n_e|0\rangle|^2 \, L_H(\varepsilon - \varepsilon_{eg} - n_e \hbar\omega_0) \tag{2}$$

In Eq. (2), we take the homogeneous lineshapes to be Lorentzian $L_H(\varepsilon) = \frac{1}{2}\Gamma_H / \left[\varepsilon^2 + \left(\frac{1}{2}\Gamma_H\right)^2\right]$ with full-width-at-half-maximum (FWHM) equal to $\Gamma_H$. The Franck-Condon overlap factors have the form $|\langle n_e|0\rangle|^2 = e^{-\lambda^2} \lambda^{2n_e}/n_e!$, where $\lambda^2 = d^2 \omega_0 / 2\hbar$ is the number of vibrational quanta absorbed by the system upon electronic excitation. $\lambda^2$ is called the Huang-Rhys parameter, and in the context of the Holstein model, it is a direct measure of the electronic-vibrational coupling strength (17).

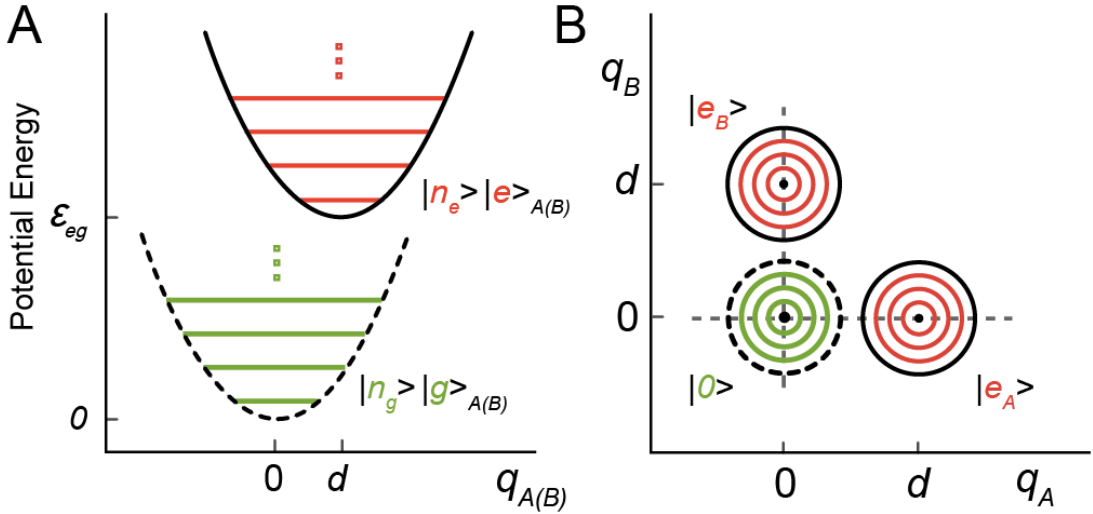

**Figure 4.** (*A*) Electronic-vibrational (vibronic) potential energy diagram for the monomer ground and excited electronic state levels, which are coupled to a single harmonic vibrational mode. (*B*) Contour diagram for the vibronic potential energy of the *AB* dimer, in the three electronic states considered in the model.

Although each monomer is chemically identical, our model includes the presence of static inhomogeneity of the transition energy $\varepsilon_{eg}$ due to variation of the local environment (45-47). We thus assign the probability that a given monomer has transition energy $\varepsilon_{eg}$ according to the Gaussian distribution $G_{I,mon}(\varepsilon_{eg}) = exp\left[-(\varepsilon_{eg} - \bar{\varepsilon}_{eg})^2/2\sigma_{I,mon}^2\right]$, which is centered at the average transition energy $\bar{\varepsilon}_{eg}$. We account for the presence of both homogeneous and inhomogeneous broadening contributions to the total line shape by using the Voigt convolution integral (47)

$$\sigma_{abs}^M(\varepsilon) = \int_{-\infty}^{\infty} \sigma_{H-abs}^M(\varepsilon - \varepsilon') \, G_{I,mon}(\varepsilon') d\varepsilon' \tag{3}$$

***Dimer Hamiltonian.*** In the collective electronic-vibrational (vibronic) basis of the *AB* dimer, we define the ground electronic state $|0\rangle = u_{n_g}^A(q_A)|g\rangle_A \otimes u_{n_g}^B(q_B)|g\rangle_B = u_{n_g n_g}(q_A, q_B)|gg\rangle$. Here we use the streamlined notation for the electronic states $|gg\rangle = |g\rangle_A|g\rangle_B$, and for the vibrational states $u_{n_g n_g}(q_A, q_B) = u_{n_g}^A(q_A) \otimes u_{n_g}^B(q_B)$. The latter emphasizes the nuclear coordinate dependence of the vibrational wave function corresponding to the product state $|n_g n_g\rangle = |n_g\rangle_A \otimes |n_g\rangle_B$. In the absence of a resonant electronic interaction, the singly-electronically-excited states are given by $|e_A\rangle = u_{n_e n_g}(q_A, q_B)|eg\rangle$ in which monomer *A* is electronically excited and monomer *B* is in the ground state, and $|e_B\rangle = u_{n_g n_e}(q_A, q_B)|ge\rangle$ in which the *A* and *B* indices are interchanged. The model potential energy surfaces corresponding to the ground and singly-electronic-excited vibronic states are illustrated in Fig. 4B.

When electronic interactions between monomers are included, the Hamiltonian of the *AB* dimer is

$$\hat{H}_{dim} = \hat{H}_A \hat{I}_B + \hat{H}_B \hat{I}_A + J\{|eg\rangle\langle ge| + |ge\rangle\langle eg|\} \hat{I}_A^{vib} \otimes \hat{I}_B^{vib} \tag{4}$$

where the expressions for the monomer Hamiltonian $\widehat{H}_M$ are given by Eq. (1). In Eq. (4), the last term describes the resonant coupling between the electronic coordinates of the two monomers. Neglecting orbital overlap, the value of $J$ is determined by the Coulomb interaction between the transition charge densities

$$J = \frac{1}{4\pi\epsilon\epsilon_0} \int_{-\infty}^{\infty} d\boldsymbol{r}_A \int_{-\infty}^{\infty} d\boldsymbol{r}_B \frac{\rho_A^{ge}(\boldsymbol{r}_A)\rho_B^{eg}(\boldsymbol{r}_B)}{|\boldsymbol{r}_A - \boldsymbol{r}_B|} \tag{5}$$

where we have defined the matrix elements $\rho_A^{ge}(\boldsymbol{r}_A) = \langle g|_A \rho(\hat{\boldsymbol{r}}_A)|e\rangle_A$ and $\rho_B^{eg}(\boldsymbol{r}_B) = \langle e|_B \rho(\hat{\boldsymbol{r}}_B)|g\rangle_B$.

In the current work, we approximate the resonant electronic coupling using the point dipole expression

$$J = \frac{|\mu_{eg}^0|^2}{4\pi\epsilon\epsilon_0} \left[ \frac{(\boldsymbol{d}_{eg}^A \cdot \boldsymbol{d}_{ge}^B)}{|\boldsymbol{R}_{AB}|^3} - 3\frac{(\boldsymbol{d}_{eg}^A \cdot \boldsymbol{R}_{AB})(\boldsymbol{R}_{AB} \cdot \boldsymbol{d}_{ge}^B)}{|\boldsymbol{R}_{AB}|^5} \right] \tag{6}$$

Here, we have defined $|\mu_{eg}^0|^2$ as the square magnitude of the monomer EDTM, and $\boldsymbol{d}_{eg}^M$ [$M = A, B$] are the unit vectors that specify each monomer direction. The point dipole approximation is justified when the inter-monomer distance is greater than two characteristic length scales: (*i*) the molecular size, and (*ii*) the transition dipole radius $|\mu_{eg}^0|/e$, where the fundamental charge unit $e = 1.60 \times 10^{-19}$ C (17). Our estimate of the molecular size is based on the output of an energy minimization calculation using the Spartan program (Wavefunction, Inc.), which suggests that the long-axis dimension of the Cy3 chromophore is ~ 14 Å. To estimate the transition dipole radius, we first determined the magnitude of the monomer EDTM ($\approx$ 12.8 D, with 1 D = 3.336 $\times$ 10$^{30}$ C m) by numerical integration of the absorption lineshape (42). We thus obtained a value for the transition dipole radius of ~ 2.7 Å. As we discuss further below, our results indicate that the smallest inter-chromophore separation under the various conditions that we studied is ~ 6 Å. While this separation is small in comparison to the molecular dimension, it remains significantly greater than the transition dipole radius. Based on this assessment alone, it is unclear how much error is introduced by the point dipole approximation. For our current purposes, we apply the point dipole approximation in order to investigate its ability to qualitatively model the resonant coupling strength and our temperature-dependent linear absorption and CD spectra.

In the presence of resonant electronic coupling, the eigenenergies and eigenstates are obtained by diagonalization of the Hamiltonian given by Eq. (4). Because of the $D_2$ symmetry of the chiral $(Cy3)_2$ dimer, the singly-electronic-excited states must be either symmetric (sign invariant, +) or anti-symmetric (sign inversion, −) under $C_2$ rotation. Spano and co-workers, who studied the redistribution of oscillator strengths within the vibronic bands of a $C_2$ symmetric dimer as a function of the resonant exchange coupling, have analyzed this problem in detail (8). They showed that the symmetric and anti-symmetric eigenstates of the coupled $AB$ dimer can be written

$$|e_{\pm}^{(\alpha)}\rangle = \sum_{n_e=0,1,\dots} \sum_{n_g=0,1,\dots} c_{\pm,n_e n_g}^{(\alpha)}(q_A, q_B)[|e_A\rangle \pm |e_B\rangle] \qquad (7)$$

In Eq. (7), $c_{\pm,n_e n_g}^{(\alpha)}$ are complex-valued coefficients that depend on the nuclear coordinates, and $\alpha$ = 0, 1, 2, … in order of increasing state energy. Note that the electronic states $|e_A\rangle$ and $|e_B\rangle$ also depend on the nuclear coordinates. We designate the transition energies of states $|e_{\pm}^{(\alpha)}\rangle$ as $\varepsilon_{\pm,\alpha}$. Moreover, $n_e$ specifies the vibrational occupancy of the electronically excited monomer site, while $n_g$ specifies that of the electronically unexcited site. It is useful to organize the singly excited states into two different categories. So-called 'one-particle states' are those with variable $n_e$ (= 0, 1, …) vibrational quanta in the shifted potential of the vibronically excited monomer, and $n_g$ = 0 quanta in the un-shifted potential of the electronically unexcited monomer. 'Two-particle states,' on the other hand, are those with variable $n_e$ in the vibronically excited monomer, and $n_g \geq 1$ in the un-shifted potential of the electronically unexcited monomer (8). The eigenstates given by Eq. (7) can thus be re-written as

$$|e_\pm^{(\alpha)}\rangle = \sum_{n_e=0,1,\ldots} c_{\pm,n_e 0}^{(\alpha)} [u_{n_e 0}|eg\rangle \pm u_{0n_e}|ge\rangle] \tag{8}$$

$$+ \sum_{n_e=0,1,\ldots} \sum_{n_g=1,2,\ldots} c_{\pm,n_e n_g}^{(\alpha)} [u_{n_e n_g}|eg\rangle \pm u_{n_g n_e}|ge\rangle]$$

where the first and second terms of Eq. (8) represent one- and two-particle contributions, respectively. For a given symmetry, the energy eigenstates are superpositions of pure-state contributions; these contributions are of like symmetry, and contain varying levels of vibrational energy.

***Dimer Absorption and CD Spectra.*** We determine the intensities of ground state accessible vibronic transitions of the *AB* dimer using the expression

$$I_\pm^{(\alpha)} = \langle 0|\boldsymbol{\mu}^{tot}|e_\pm^{(\alpha)}\rangle \langle e_\pm^{(\alpha)}|\boldsymbol{\mu}^{tot}|0\rangle \tag{9}$$

where the collective EDTM is given by $\boldsymbol{\mu}^{tot} = \boldsymbol{\mu}_{eg}^A + \boldsymbol{\mu}_{eg}^B$. The absorption spectrum of the *AB* dimer may thus be decomposed into symmetric and anti-symmetric transition manifolds, which are polarized along the directions of the molecular frame *x*- and *y*-axes, respectively (see Fig. 3 for coordinate system definitions).

$$\sigma_{H-abs}^{dim}(\varepsilon) = \sigma_{H-abs,+}^{dim}(\varepsilon) + \sigma_{H-abs,-}^{dim}(\varepsilon) \tag{10}$$

with

$$\sigma_{H-abs,\pm}^{dim}(\varepsilon) = \sum_{\alpha} \left|\left\langle 0 | \boldsymbol{\mu}^{tot} | e_{\pm}^{(\alpha)} \right\rangle\right|^2 L_H(\varepsilon - \varepsilon_{\pm,\alpha}) \qquad (11)$$

The CD spectrum is similarly decomposed into polarized components

$$CD_H^{dim}(\varepsilon) = \sum_{\alpha} RS_{H+}^{(\alpha)} L_H(\varepsilon - \varepsilon_{+,\alpha}) + \sum_{\alpha} RS_{H-}^{(\alpha)} L_H(\varepsilon - \varepsilon_{-,\alpha}) \qquad (12)$$

where the rotational strengths for the symmetric and anti-symmetric transitions are given by

$$RS_{H\pm}^{(\alpha)} = \frac{\varepsilon_{eg}}{4\hbar c |\mu_{eg}^0|^2} \left\langle 0 | \boldsymbol{\mu}^A | e_{\pm}^{(\alpha)} \right\rangle \times \left\langle e_{\pm}^{(\alpha)} | \boldsymbol{\mu}^B | 0 \right\rangle \cdot \boldsymbol{R}_{AB} \qquad (13)$$

For both the absorption and CD of the *AB* dimer, we take into account the effects of inhomogeneous broadening using a pseudo-Voigt profile (48) that approximates the convolution given by Eq. (3), except using the Gaussian distribution $G_{I,dim}(\varepsilon_{eg})$ specific to the dimer.

Spano and co-workers performed a systematic analysis of the effects of varying exciton interaction strength on the polarized components $(+/-)$ of the absorption and CD spectra of a chiral $C_2$ symmetric dimer (8). They used a perturbation theoretical approach to derive expressions for the oscillator strengths of the first two vibronic lineshapes in the weak exciton-coupling regime (i.e. for $|J_{\pm}| \ll \lambda^2 \hbar \omega_0$, where $J_{\pm} = \pm J$). In the weak-coupling regime, only single-particle contributions to the eigenstates described by Eq. (8) need be considered, so that each $v_t - 0$ vibronic band of the monomer spectrum is split into just one upper (+) and one lower (−) energy component. In this regime, the Davydov splitting for each vibronic band of the *AB* dimer is approximately equal to $2Je^{-\lambda^2}\lambda^{2v_t}/v_t!$. The ratio of the $0 - 0$ to $1 - 0$ line strengths is given by

$$\frac{I_{\pm}^{(0-0)}}{I_{\pm}^{(1-0)}} = \frac{1}{\lambda^2}\left[\frac{1-G(0;\lambda^2)e^{-\lambda^2}J_{\pm}/\hbar\omega_0}{1-G(1;\lambda^2)e^{-\lambda^2}J_{\pm}/\hbar\omega_0}\right]^2 \qquad (14)$$

where the effect of the electronic-vibrational coupling is described by the function

$$G(\nu_t;\lambda^2) = \sum_{\substack{u=0,1,\ldots \\ (u\neq \nu_t)}} \frac{\lambda^{2u}}{u!\,(u-\nu_t)} \qquad (15)$$

In Eq. (15), the index $\nu_t$ (= 0,1,…) designates the vibronic band of the absorptive transition. As we shall show below, Eq. (14) captures the essential behavior of the polarized components of the absorption and CD for the (Cy3)$_2$ DNA construct.

From our analysis of the monomer spectra over the pre-melting range 15 – 60 °C, we see that the intramolecular parameters are approximately independent of temperature. We thus obtain the average values: $\lambda^2$ = 0.55, $\hbar\omega_0$ = 1,115 cm$^{-1}$, and $\varepsilon_{eg}$ = 18,276 cm$^{-1}$. Substituting these values into Eqs. (14) and (15), we obtain $G(\nu_t = 0; 0.55)e^{-0.55} = +0.367$, $G(\nu_t = 1; 0.55)e^{-0.55} = -0.481$, and

$$\frac{I_{\pm}^{(0-0)}}{I_{\pm}^{(1-0)}} = \frac{1}{\lambda^2}\left[\frac{1-0.367 J_{\pm}/\hbar\omega_0}{1+0.481 J_{\pm}/\hbar\omega_0}\right]^2 \qquad (16)$$

From Eq. (16), we see that the expected effect of increasing exciton interaction $J$ is to decrease the ratio of the symmetrically polarized components $I_+^{(0-0)}/I_+^{(1-0)}$, and to increase the ratio of the anti-symmetrically polarized components $I_-^{(0-0)}/I_-^{(1-0)}$. Thus, the upper energy symmetric excitons, which are polarized in the direction of $\boldsymbol{\mu}_{eg}^A + \boldsymbol{\mu}_{eg}^B$ (the $x$-axis), tend to be more heavily weighted by single-particle contributions with higher vibrational quantum number ($n_e \geq 1$) than do the lower energy excitons, which are polarized in the direction of $\boldsymbol{\mu}_{eg}^A - \boldsymbol{\mu}_{eg}^B$ (the $y$-axis). We expect this effect to become more pronounced with increasing resonant coupling interaction.

***Numerical Calculations.*** To perform numerical calculations for the Holstein model of the *AB* dimer, it is convenient to transform Eq. (4) to the energy basis using excitation creation / annihilation operators (17). We adopt the bosonic operators $\hat{b}_M^\dagger = \sqrt{1/2}\left[\sqrt{m\omega_0/\hbar}\,\hat{q}_M - i\hat{p}_M/\sqrt{m\omega_0\hbar}\right]$ and $\hat{b}_M = \sqrt{1/2}\left[\sqrt{m\omega_0/\hbar}\,\hat{q}_M + i\hat{p}_M/\sqrt{m\omega_0\hbar}\right]$ for the creation and annihilation of vibrational quanta, respectively, within the harmonic potential surfaces associated with the ground electronic states. These operators obey the boson commutation relation $[\hat{b}_{M'}, \hat{b}_M^\dagger] = \delta_{M'M}$, where $\delta_{M'M}$ is the Kronecker delta function and $M', M \in \{A, B\}$. We further define the operators $\hat{c}_M^\dagger$ and $\hat{c}_M$ to represent the creation and annihilation of electronic quanta, respectively, within the two-level monomer $M = A, B$. Because a single two-electronic-level molecule cannot be excited twice, these operators obey the fermion commutation relation $[\hat{c}_{M'}, \hat{c}_M^\dagger] = \delta_{M'M}(1 - 2\hat{c}_M^\dagger \hat{c}_M)$. Using the above definitions and the expression for the Huang-Rhys parameter $\lambda^2 = d^2\omega_0/2\hbar$, the monomer Hamiltonian [Eq. (1)] can be recast as

$$\hat{H}_M = \{\varepsilon_{eg}\hat{c}_M^\dagger \hat{c}_M + \hbar\omega_0 \hat{b}_M^\dagger \hat{b}_M \hat{I}_M^{elec} + \hbar\omega_0 \hat{c}_M^\dagger \hat{c}_M [\lambda(\hat{b}_M^\dagger + \hat{b}_M) + \lambda^2]\}\hat{I}_{M' \neq M} \tag{17}$$

where, as before, the monomer index $M = A, B$. In Eq. (17), the first term on the right-hand side describes the electronic energy of the system, the second term describes the vibrational energies, and the final term describes the coupling between electronic and vibrational states within each monomer. The Hamiltonian for the coupled *AB* system [Eq. (4)] can be rewritten as

$$\hat{H}_{dim} = \hat{H}_A \hat{I}_B + \hat{H}_B \hat{I}_A + J(\hat{c}_A^\dagger \hat{c}_B + \hat{c}_A \hat{c}_B^\dagger)\,\hat{I}_A^{vib} \otimes \hat{I}_B^{vib} \tag{18}$$

where the expressions for the monomer Hamiltonian $\hat{H}_{A(B)}$ are given by Eq. (17).

***Multi-Parameter Optimization Procedure.*** In order to characterize the absorption and CD spectra of the Cy3 monomer and (Cy3)$_2$ dimer DNA constructs [given by Eqs. (2), (11) and (12)], it was necessary to obtain an estimate of the homogeneous line width. As we discuss further below, the effects of pure dephasing – i.e. coupling to the phonon bath that rapidly modulates the monomer transition energy – dominate the homogeneous line width (47). Since pure dephasing exhibits only a weak temperature-dependence, the homogeneous line width is not expected to change significantly over the range of temperatures we investigated (15 – 85 °C). We therefore used 2DFS to determine the FWHM Lorentzian line width, $\Gamma_H$ = 186 cm$^{-1}$, of both monomer and dimer-labeled DNA samples at room temperature, and we assumed this value to be constant for our analysis.

We see from Eqs. (1) – (3) that absorption spectra of the Cy3 monomer may be characterized using four independent parameters: (i) the monomer electronic transition energy $\varepsilon_{eg}$, (ii) the Huang-Rhys electronic-vibrational coupling parameter $\lambda^2$, (iii) the single-mode vibrational frequency $\omega_0$, and (iv) the spectral inhomogeneity parameter of the monomer specified by the Gaussian standard deviation $\sigma_{I,mon}$. It is also necessary to specify the magnitude of the monomer EDTM $|\mu_{eg}^0|$ = 12.8 D [see Eq. (2)], which we determined by integrating the experimental absorption lineshape as in past work (42). To characterize fully the absorption and CD spectra of the coupled (Cy3)$_2$ dimer [see Eqs. (4), (11), (12)], we must additionally specify: v) the inter-chromophore separation $R_{AB}$, (vi) the inter-chromophore twist angle $\phi_{AB}$, and (vii) the spectral inhomogeneity parameter $\sigma_{I,dim}$ associated with the dimer. The values of the structural parameters $R_{AB}$ and $\phi_{AB}$ determine the resonant coupling strength $J$ according to Eq. (6).

In order to obtain the most favorable comparison between simulated and experimental absorption and CD spectra, we implemented an automated multi-variable regression analysis to efficiently explore the space of input parameters (i) – (vii). The procedure is similar to one we have used in past studies (42, 49, 50), in which a random search algorithm generates an initial set of input parameters, and commercial software (KNITRO) (51) is used to refine the corresponding solutions. For each set of input trial parameters, we calculate a linear least-squares target function $\chi^2$, which guides the selection of parameter values for subsequent iterations. The optimized solutions correspond to minimization of the target function.

We initially applied the above optimization procedure to the Cy3 monomer absorption spectrum taken at 15 °C by minimizing the target function $\chi^2_{abs,mon}(\varepsilon_{eg}, \lambda^2, \hbar\omega_0, \sigma_{I,mon})$. We next applied the optimization procedure to the monomer data sets taken at each temperature. We thus determined optimized values of the parameters (i) – (iv) as a function of temperature, which are listed in Table SI of the SI. We note that the parameters $\varepsilon_{eg}$, $\lambda^2$, and $\omega_0$ did not appear to depend on temperature, while the monomer inhomogeneity parameter $\sigma_{I,mon}$ increased with temperature.

We next performed joint optimizations on the (Cy3)$_2$ dimer absorption and CD spectra. For each temperature, we used the optimized values of the monomer parameters (i) – (iii) as inputs to the dimer calculations. From our convergence tests using the 15 °C data, we concluded that occupation of at least six vibrational levels, in both the monomer ground and excited electronic-state manifolds, were needed to obtain converged simulations. We thus determined the three remaining trial function parameters [(v) – (vii)] by minimizing the target function:

$$\chi^2_{tot}(R_{AB}, \phi_{AB}, \sigma_{I,dim}) = \chi^2_{abs,dim}(R_{AB}, \phi_{AB}, \sigma_{I,dim}) + \chi^2_{CD,dim}(R_{AB}, \phi_{AB}, \sigma_{I,dim}) \quad (19)$$

Error bars associated with the optimized parameters were determined by a 1% deviation of the target function from its minimized value. The results of our optimization analysis of the dimer spectra are presented in Table II, and discussed further below.

## IV. Discussion of Results

***Estimation of the homogeneous line widths of the Cy3 monomer and (Cy3)$_2$ dimer-labeled DNA constructs.*** In Fig. 5, we present 2DFS measurements of the monomer and dimer labeled DNA samples at room temperature. For these measurements, we tuned the laser center wavelength across the low energy 0 – 0 and 1 – 0 sub-bands of the absorption spectra (515 – 550 nm). The rephasing 2DFS spectra exhibited quasi-elliptical 2D lineshapes, with representative examples shown in Fig. 5. Rephasing 2DFS measurements have the property that inhomogeneous line broadening does not contribute to the 2D spectrum along the anti-diagonal direction (45). We thus determined the homogeneous line width from the anti-diagonal cross-sectional width. We compared fits of the diagonal and anti-diagonal cross-sections of the 2D spectra using both Lorentzian and Gaussian functions [see Figs. 5C, 5D, 5G and 5H]. While the diagonal cross-sectional width of the 2D spectrum (FWHM, 505 cm$^{-1}$) closely matched that of the laser bandwidth (555 cm$^{-1}$), we found that the anti-diagonal cross-sectional width varied only slightly with laser center wavelength. We thus determined the average value of the Lorentzian FWHM $\Gamma_H = 186$ cm$^{-1}$, which corresponds to the total dephasing time $T_2 = (\pi c \Gamma_H)^{-1} \cong 57$ fs.

The total dephasing time is related to the population relaxation time ($T_1$) and the pure dephasing time ($T_2'$) according to $(T_2)^{-1} = (2T_1)^{-1} + (T_2')^{-1}$ (47). For the Cy3 DNA constructs, the value of $T_1$ can be estimated using the room temperature fluorescence lifetime $\tau_f \sim$ 162 ps (28). It is known that the fluorescence lifetime of Cy3 DNA constructs can vary with temperature due to the thermal activation of intramolecular photo-isomerization processes (28, 52-54). Nevertheless, such picosecond processes are orders of magnitude slower than those of pure dephasing, which dominate the homogeneous line width. As mentioned above, pure dephasing results from rapid fluctuations of the electronic transition energy due to interactions with the phonon bath. In proteins and disordered media, the pure dephasing time typically follows the relatively weak, power law temperature-dependence $T_2' \sim T^{1.3}$ (47, 55). This suggests that the homogeneous line width is relatively insensitive to temperature. We therefore used the above determined value $\Gamma_H = 186$ cm$^{-1}$ for our analyses of the linear absorption and CD spectra, as discussed below.

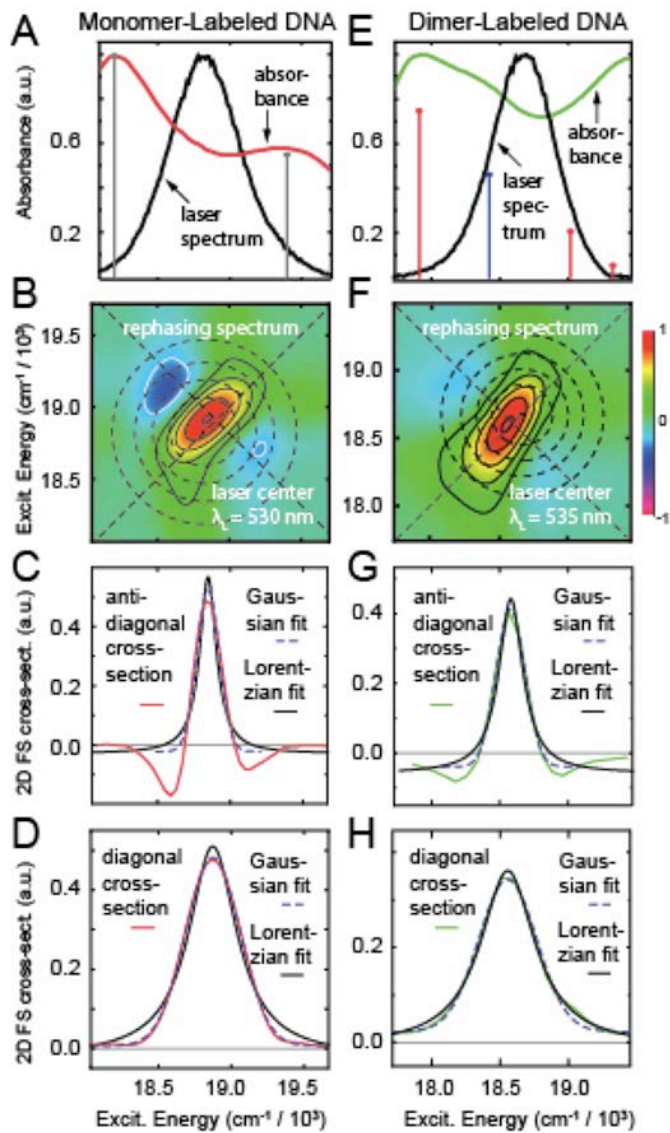

**Figure 5.** 2DFS rephasing spectra of the Cy3 monomer (*A – D*) and the (Cy3)$_2$ dimer-labeled DNA constructs (*E – H*). (*A & E*). Spectral overlap of the absorbance and laser excitation. (*B & F*) 2DFS rephasing spectra of the monomer and dimer construct. The concentric circles indicate the laser spectral overlap. The diagonal and anti-diagonal 2D cross-sections are marked with dashed lines. (*C & G*) Anti-diagonal lineshape of the monomer and dimer fit to Lorentzian and Gaussian functions, respectively. (*D & H*) Diagonal lineshape of the monomer and dimer fit to Lorentzian and Gaussian functions, respectively.

***Absorbance and CD Spectra.*** We studied the temperature-dependence of the absorption and CD spectra of both the Cy3 monomer and the (Cy3)$_2$ dimer-labeled DNA constructs. As described in the previous sections, we found that the monomer absorption spectrum did not vary significantly with temperature (see Table SI and Fig. S2 of the SI). In Fig. 6, we present absorption and CD spectra for the dimer at representative temperatures overlaid with optimized simulations of the polarized symmetric (+) and anti-symmetric (−) components. The agreement between experiment and theory is very good over the full range of temperatures we investigated. In Table II, we list as a function of temperature the output values of our optimization procedure, which include the resonant coupling strength $J$, the inter-chromophore twist angle $\phi_{AB}$, the inter-chromophore separation $R_{AB}$, and the spectral inhomogeneity parameter $\sigma_{I,dim}$. For temperatures above the melting transition (at 65°C), the absorption spectrum of the dimer became indistinguishable from that of the Cy3 monomer DNA construct, signifying the complete separation between the conjugated single DNA strands.

We first consider the absorption and CD spectrum of the (Cy3)$_2$ dimer DNA construct at 15°C, which is the lowest temperature we investigated. We obtained optimized values for the structural parameters $J = 529$ cm$^{-1}$, $\phi_{AB} = 82.9°$, $R_{AB} = 5.8$ Å, and $\sigma_{I,dim} = 292$ cm$^{-1}$. The values for $R_{AB}$ and $\phi_{AB}$ are consistent with the local conformation of the (Cy3)$_2$ dimer depicted in Fig. 1, which shows the two Cy3 monomers positioned closely within the DNA duplex with nearly orthogonal relative orientation. The magnitude of the resonant coupling strength $J$ is greater than the spectral inhomogeneity $\sigma_{I,dim}$ of the system, which is a necessary condition for the dimer to support delocalized excitons. Furthermore, because the coupling strength is comparable to the intramolecular vibrational relaxation energy (i.e. $J \sim \lambda^2 \hbar\omega_0 = 602$ cm$^{-1}$, where we have used $\lambda^2 = 0.54$ and $\hbar\omega_0 = 1,116$ cm$^{-1}$), the dimer must exist in the intermediate-to-strong exciton-coupling regime. From these observations, we conclude that at 15 °C, the perturbation theory description of the exciton band structure [summarized by Eqs. (14) – (16)] should not strictly hold.

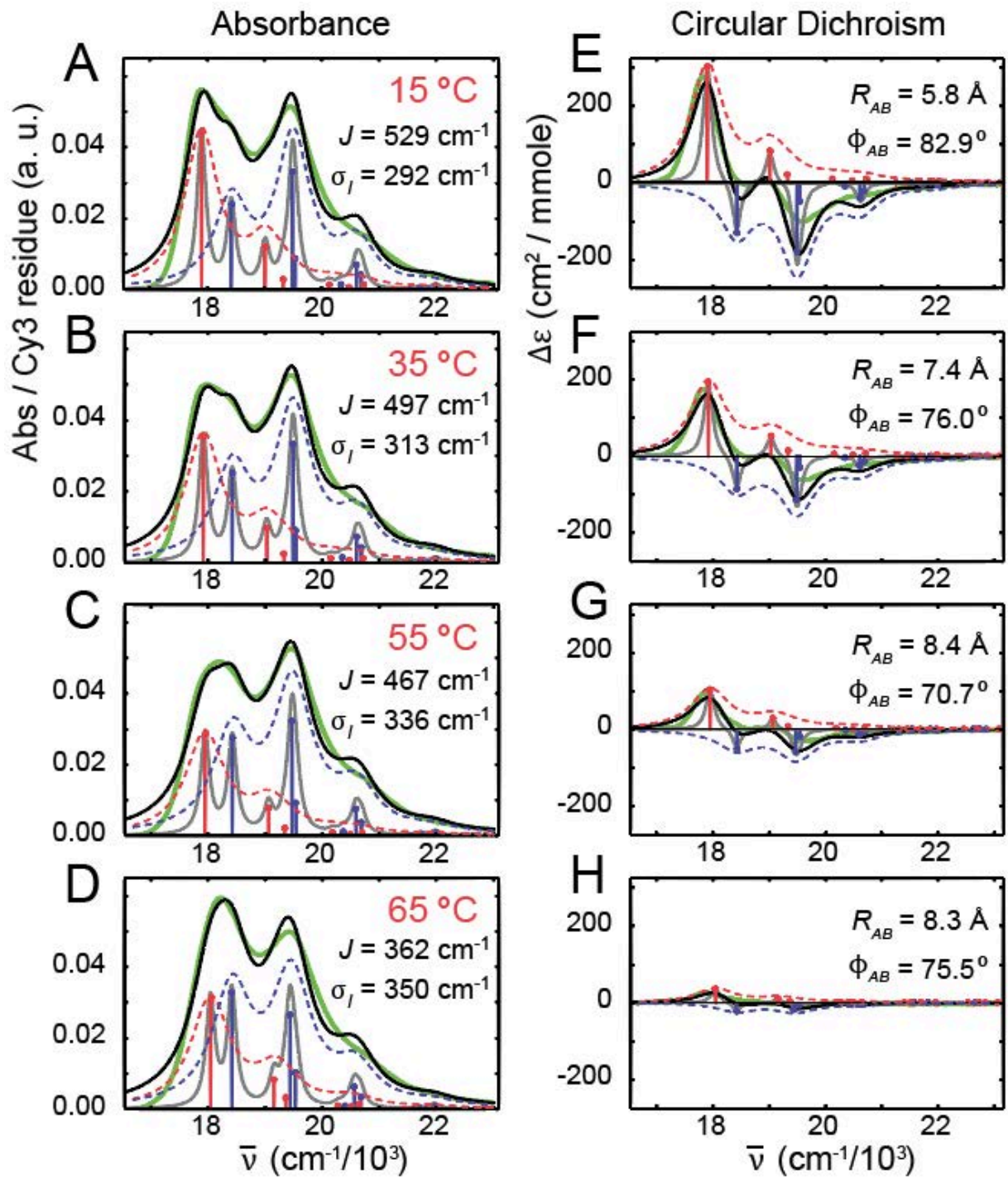

**Figure 6.** Temperature-dependent absorption (*A-D*) and CD spectra (*E-H*) for $(Cy3)_2$ dimer labeled DNA constructs. Experimental spectra are shown in solid green, the model homogeneous lineshapes in solid gray, and the model total lineshapes (inhomogeneous-plus-homogeneous) in solid black. Symmetric and anti-symmetric transitions determined from the model are shown as blue and red sticks, respectively. Symmetric and anti-symmetric contributions to the inhomogeneous lineshapes are shown as dashed blue and red curves, respectively.

From our simulated model fits to the absorption and CD spectra, we determined the experimental Davydov splitting of the 0 – 0 vibronic band. We define the Davydov splitting for the 0 – 0 band as the energy difference between the single upper energy (symmetric, +) exciton and the single lower energy (anti-symmetric, –) exciton within this band. Our analysis of the 15 °C spectra reveal a pronounced splitting ($DS_{0-0}$ = 532 cm$^{-1}$), with the lower energy anti-symmetric exciton exhibiting greater intensity than that of the upper energy symmetric exciton (see Figs. 6A and 6E). We note that the pronounced bisignate splittings of individual vibronic bands of the CD spectrum can be understood from the opposite sign contributions of the symmetric and anti-symmetric excitons [see Eqs. (12) and (13)] (8). This condition follows from the chiral $D_2$ symmetry of the coupled (Cy3)$_2$ dimer, which leads to significant oscillator strength contributions from both symmetric and anti-symmetric excitons.

**Table II.** Optimized values of the structural parameters of the (Cy3)$_2$ dimer DNA construct at various temperatures, obtained from the Holstein model fit to absorption and circular dichroism spectra. These calculations used as inputs the electric transition dipole moment (EDTM) $|\mu_{eg}^0|$ = 12.8 D, and for each temperature, the electronic transition energy $\varepsilon_{eg}$, the vibrational mode frequency $\omega_0$, and the Huang-Rhys parameter $\lambda^2$ obtained from our analyses of the absorption spectra of the Cy3 monomer DNA construct (see Table SI of the SI). The parameters listed are the resonant coupling strength $J$, the inter-chromophore twist angle $\phi_{AB}$, the inter-chromophore distance $R_{AB}$, and the standard deviation of the Gaussian inhomogeneous disorder function $\sigma_{I,dim}$. Structural parameters are presented at temperatures below the melting transition at 65 °C, for which the dimer model may reasonably be applied. Error bars were calculated based on a 1% deviation of the target function from its optimized value.

| $T$ (°C) | $J$ (cm$^{-1}$) | $\phi_{AB}$ (°) | $R_{AB}$ (Å) | $\sigma_{I,dim}$ (cm$^{-1}$) |
|---|---|---|---|---|
| 15 | 529 +51/-129 | 82.9 +0.9/−0.4 | 5.8 +0.3/−0.1 | 292 +24/−6 |
| 25 | 514 +63/-120 | 80.1 +1.3/-0.5 | 6.5 +0.3/-0.1 | 302 +23/-6 |
| 35 | 497 +61/-114 | 76.0 +1.7/-0.8 | 7.4 +0.3/-0.2 | 313 +20/-7 |
| 45 | 483 +80/-99 | 72.3 +2.0/-1.3 | 8.0 +0.3/-0.2 | 325 +18/-8 |
| 55 | 467 +64/-94 | 70.7 +1.9/-1.4 | 8.4 +0.3/-0.2 | 336 +15/-9 |
| 60 | 449 +75/-82 | 70.4 +1.9/-1.6 | 8.5 +0.3/-0.2 | 345 +14/-10 |
| 65 | 362 +83/-82 | 75.5 +1.7/-1.6 | 8.3 +0.4/-0.3 | 350 +13/-12 |

We determined the Davydov splitting of the 1 – 0 vibronic band ($DS_{1-0} = 327$ cm$^{-1}$) by considering only single-particle exciton contributions, which are expected to dominate the absorption and CD spectra in the limit of weak exciton coupling. In this limit, one upper energy symmetric state and one lower energy anti-symmetric state are much larger in magnitude than the remaining states within 1 – 0 vibronic band. This appears to be valid for the 15 °C sample, as well as for samples at higher temperatures (see below). We note that the above values for $DS_{0-0}$ and $DS_{1-0}$ are similar in magnitude to the theoretical prediction (616 cm$^{-1}$ and 333 cm$^{-1}$, respectively) given by the factor $2Je^{-\lambda^2}\lambda^{2v_t}/v_t!$. Based on the simulated spectra, we determined the ratios of the vibronic band intensities $I_+^{(0-0)}/I_+^{(1-0)} = 0.62$ for the symmetric exciton, and $I_-^{(0-0)}/I_-^{(1-0)} = 2.60$ for the anti-symmetric exciton. We found that the symmetric (anti-symmetric) band intensity ratio increases (decreases) in the coupled dimer relative to that of the free monomer [$I_{mon}^{(0-0)}/I_{mon}^{(1-0)} = 1.60$], as suggested by the perturbation theory (8). However, these values are considerably smaller in magnitude than those predicted (0.84 and 4.31, respectively) by Eq. (16). These findings further support that at 15 °C, the system resides in the intermediate-to-strong exciton-coupling regime. While structural disorder is significant, it does not mask the effects of the intermediate-to-strong exciton delocalization.

As the temperature was increased over the range 15 – 65 °C, the effects of exciton coupling on the dimer absorption and CD spectra became less pronounced (see Fig. 6). The Davydov splitting for both the 0 – 0 and 1 – 0 vibronic bands decreased continuously, as did the finite amplitudes of the CD signal. We note that the agreement between experimental and theoretical values of $DS_{0-0}$ and $DS_{1-0}$ is good over the full range of temperatures (see Table SII). Moreover, the vibronic band intensity ratio $I_{+(-)}^{(0-0)}/I_{+(-)}^{(1-0)}$ of the symmetric (anti-symmetric) exciton appeared to increase (decrease) with increasing temperature. Comparison between experimental and theoretical values for $I_{+(-)}^{(0-0)}/I_{+(-)}^{(1-0)}$ (see Table SIII) appears to become more favorable at elevated temperatures, consistent with the system undergoing a transition to the weak exciton-coupling regime.

We see that the temperature-dependent properties of the (Cy3)$_2$ dimer DNA construct are correlated to a systematic change in the resonant coupling strength $J$ between the Cy3 monomer subunits. This is due to the temperature sensitivity of cooperative interactions between constituent nucleobases (e.g. base stacking interactions, Watson-Crick hydrogen bonding, etc.), which stabilize the right-handed helical structure of the DNA duplex. The temperature-dependent disruption of local DNA secondary structure is reflected by systematic changes in the conformation of the (Cy3)$_2$ dimer, which are characterized by the structural parameters listed in Table II and plotted in Fig. 7.

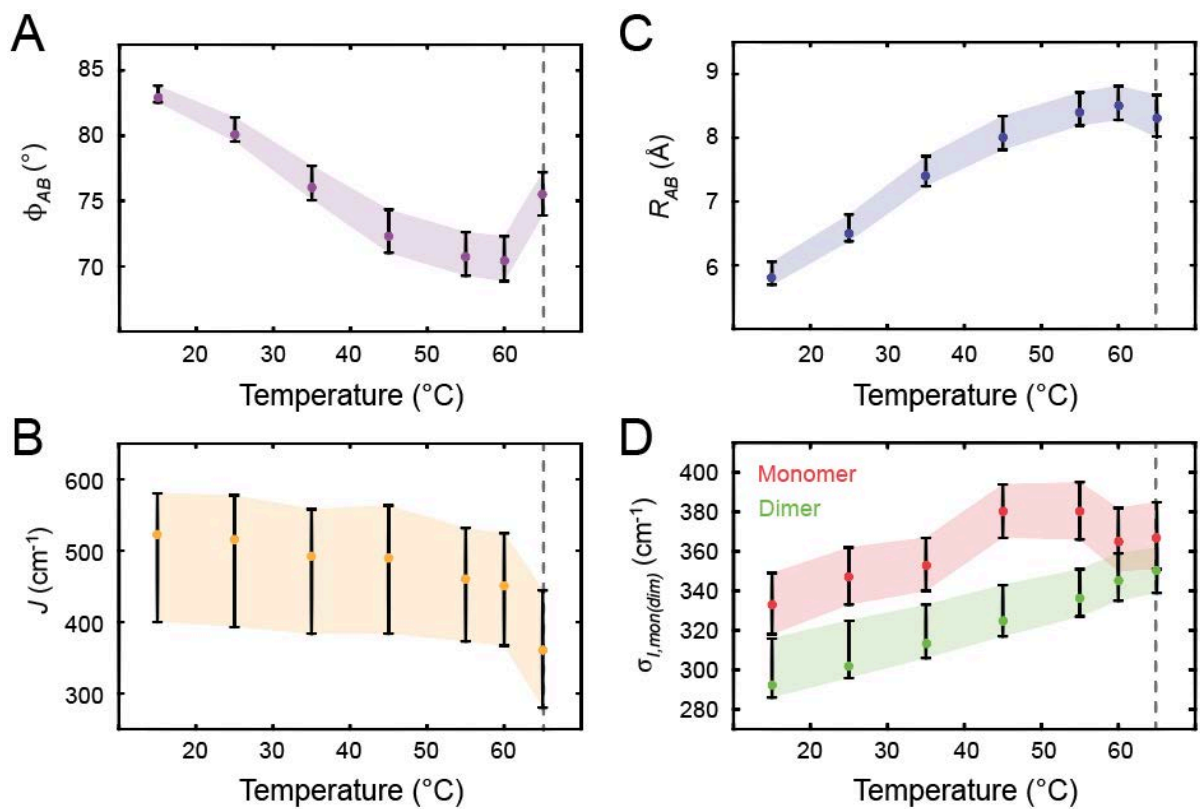

**Figure 7.** Temperature-dependent optimized parameters from (Cy3)$_2$ dimer absorption and CD spectra. Error bars were calculated based on a 1% deviation of the target function from its optimized value. The dashed line at 65 °C indicates the melting transition temperature $T_m$ of the DNA constructs. **(A)** Inter-chromophore twist angle; **(B)** Resonant electronic coupling parameter; **(C)** Inter-chromophore separation; and **(D)** Spectral inhomogeneity parameter associated with the Cy3 monomer and the (Cy3)$_2$ dimer DNA constructs.

As shown in Fig. 7, the structural parameters of the (Cy3)$_2$ dimer vary continuously over the range of temperatures 15 – 60 °C. The inter-chromophore separation $R_{AB}$ increases from 5.8 – 8.5 Å, the inter-chromophore twist angle $\phi_{AB}$ decreases from 82.9 – 70.4°, the resonant coupling $J$ decreases from 529 – 449 cm$^{-1}$, and the spectral inhomogeneity parameter $\sigma_{I,dim}$ increases from 292 – 345 cm$^{-1}$. The spectral inhomogeneity is a measure of the disorder of the local DNA environment experienced by the chromophores. In Fig. 7D, we compare the spectral inhomogeneity for both the Cy3 monomer and the (Cy3)$_2$ dimer DNA constructs as a function of temperature. The values of both the parameters $\sigma_{I,mon}$ and $\sigma_{I,dim}$ increase with temperature, which suggests the presence in both species of a broad distribution of thermally populated sub-states. It is interesting that the level of static disorder appears to be greater in the Cy3 monomer DNA construct in comparison to that of the (Cy3)$_2$ dimer construct. The disorder parameter of the monomer $\sigma_{I,mon}$ increases monotonically with temperature over the range 15 – 45 °C, and then undergoes a gradual decrease over the range 45 – 65 °C to the same value as that of the dimer $\sigma_{I,dim}$ at the melting transition. This is likely a reflection of the less favorable packing conditions of the Cy3 monomer DNA construct, for which a single thymine base is positioned across from the Cy3 chromophore on the opposing DNA single-strand. Presumably, whatever structural constraints restrict the conformational space of the Cy3 monomer DNA construct are relaxed at temperatures in excess of 45 °C.

The DNA construct undergoes a denaturation, or 'melting' transition at 65 °C, as evidenced by an increased optical absorption of the nucleobases at 260 nm (see Fig. S4). Immediately below the melting transition at 65 °C, the absorption spectrum develops an abrupt increase in the intensity of the 0 – 0 vibronic band and a concomitant decrease in the intensity of the 1 – 0 band, resulting in a spectrum nearly identical to that of the Cy3 monomer DNA construct (compare Fig. 6A to 6D). Thus, when the DNA strands dissociate at the melting transition, the electronic properties become those of the isolated monomers due to the complete disruption of the resonant coupling. At temperatures near and above 65 °C, the Holstein dimer model cannot accurately represent the electronic properties of the system because the electronic interaction is disrupted by the DNA strand separation. This leads to increased error bars associated with the optimized conformational parameters above 65 °C.

## V. Conclusions

In this work, we studied the absorption and CD spectra of a (Cy3)$_2$ dimer, which was rigidly positioned within the sugar-phosphate backbone of double-stranded DNA. We applied an essential-state Holstein model to characterize the temperature-dependent excitons supported by the dimer, for which the electronic and vibrational states of the isolated monomers are internally coupled. At the lowest temperature we studied (15°C), the system exhibited intermediate-to-strong resonant coupling ($J \sim 500$ cm$^{-1}$), comparable in magnitude to the vibrational relaxation energy of the constituent monomers ($\lambda^2 \hbar \omega_0 \sim 600$ cm$^{-1}$). Under these conditions, the dimer can support delocalized excitons composed of symmetric and anti-symmetric superpositions of electronic-vibrational product states. This electronic structure is a consequence of the co-facial geometry of the (Cy3)$_2$ dimer with inter-chromophore twist angle $\phi_{AB} \sim 80°$ and inter-chromophore separation $R_{AB} \sim 6$ Å (Fig 1).

As the temperature was increased towards the ds – ss DNA melting temperature ($T_m = 65$ °C), the resonant coupling strength gradually decreased over an $\sim 80$ cm$^{-1}$ range, while the Hamiltonian parameters characteristic of the monomer (i.e. the transition energy $\varepsilon_{eg}$, the Huang-Rhys electronic-vibrational coupling parameter $\lambda^2$, and the vibrational frequency $\omega_0$) remained approximately independent of temperature. This is a consequence of the sensitivity of the local secondary structure of the dsDNA to temperature, which affects the inter-chromophore separation and twist angle, but not the electronic-vibrational properties internal to each monomer. Our accompanying 2DFS measurements allowed us to estimate the spectral homogeneous line width, which was approximately the same for both the monomer and dimer (FWHM $\Gamma_H = 186$ cm$^{-1}$, corresponding to coherence time $\tau_c = (\pi c \Gamma_H)^{-1} \cong 57$ fs). The spectral inhomogeneity parameters of the monomer and dimer (given by the standard deviations $\sigma_{I,mon}$ and $\sigma_{I,dim}$, respectively) exhibited a systematic increase with temperature, signifying that the probe chromophores experience locally disordered, thermally activated regions of the DNA duplex, well below the melting transition. While the magnitude of spectral inhomogeneity is significant across the 15 – 65°C temperature range (290 – 350 cm$^{-1}$), the effects of exciton delocalization within the (Cy3)$_2$ dimer are not dominated by the spectral inhomogeneity.

Although the Holstein model for the exciton-coupled (Cy3)$_2$ dimer is relatively simple as it assumes a single internal vibrational mode for each monomer, the model appears to capture the essential features of the experimental absorption and CD spectra over the full range of temperatures we investigated. The success of the Holstein model may be due in large part to the presence of an intense Raman-active vibration at $\sim 1,200$ cm$^{-1}$, which is attributed primarily to symmetric stretching of the trimethine bridge of the Cy3 chromophore (56). Previous work by others have examined similar systems, such as dsDNA supported (Cy5)$_2$ dimers (57), and (Cy3)$_2$ dimers attached to DNA using flexible linkers (58). However, those studies did not account for the influence of the vibrational states of the cyanine chromophores, which led to claims of solely J-type and H-type dimer conformations, respectively. On the other hand, a similar Holstein model was previously used to describe a synthetically derived (Cy3)$_2$ dimer, which was rigidly held to a single achiral conformation (i.e., a racemic mixture) using covalent aliphatic groups (11, 12). While the intramolecular parameters and transition dipole moment for that system were roughly the same as those we found for the (Cy3)$_2$ DNA dimer of the current work, the electronic properties of the synthetic (Cy3)$_2$ dimer are notably different. The intermolecular structural parameters $\phi_{AB}$ = 18° and $R_{AB}$ = 10 Å correspond to a significantly stronger resonant coupling strength $J$ = 820 cm$^{-1}$ than the value we obtained for the most structured (Cy3)$_2$ DNA conformation at 15 °C, and the electronic properties are dominated by the H-type (symmetric) exciton. It is a consequence of the relatively large inter-chromophore twist angle $\phi_{AB}$ = 83° of the $D_2$ symmetric chiral conformation of the (Cy3)$_2$ DNA system that both H- (symmetric) and J- (anti-symmetric) type exciton components contribute significantly to the absorption and CD spectra.

For the (Cy3)$_2$ DNA system, temperature variation allows the resonant coupling strength to be 'tuned' across the intermediate-to-strong exciton-coupling regime, while the Hamiltonian parameters characterizing the internal properties of the Cy3 monomers are approximately constant. Moreover, spectral inhomogeneity (i.e. local site-energy disorder) is significant in this system, and is likely due to the presence of local structural fluctuations of the DNA backbone and base stacking that influence the packing of the chromophore probes. Such local fluctuations of DNA are termed DNA 'breathing,' and are thought to be significant to molecular biological processes such as protein-DNA binding and protein function (36). The above properties of the (Cy3)$_2$ dimer DNA construct suggest that it may be employed as a useful model system to test fundamental concepts of protein-DNA interactions, and the role of electronic-vibrational coherence in electronic energy migration within exciton-coupled bio-molecular arrays.

**Supplementary Material**

Supplementary material contains temperature-dependent optimized parameters of the Holstein model fit to the absorption and CD spectra of Cy3 (monomer and dimer) DNA constructs. Davydov splittings are provided for both the $0-0$ and $1-0$ vibronic bands, and the vibronic band intensity ratios $I_{\pm}^{(0-0)}/I_{\pm}^{(1-0)}$ for the symmetric and anti-symmetric excitons. A comparison is shown between optimized fits to absorption and CD spectra using a Gaussian versus Lorentzian homogeneous line shape. Temperature-dependent UV spectra of the Cy3 DNA constructs are presented, which establish the denaturation temperature. Ball-and-stick and space-filling structural models are provided for visualization of the (Cy3)$_2$ dimer DNA construct.

**Acknowledgements**

We thank Prof. Peter H. von Hippel and members of the von Hippel group for useful discussions pertaining to DNA breathing and other factors that affect DNA stability. We also thank Prof. Frank Spano for useful discussions about H- and J-coupling in exciton-coupled molecular dimers. This work was supported by the John Templeton Foundation (RQ-35859 to A.H.M. M.R., and A.A.-G. as co-PIs), by the National Science Foundation Chemistry of Life Processes Program (CHE-1608915 to A.H.M.), and by the National Institutes of General Medical Sciences (NIGMS Grant GM-15792 to A.H.M. as a co-PI). L.K. acknowledges support as a Rosaria Haugland Graduate Research Fellow. A.A.-G. and N.S. acknowledge the Center for Excitonics, an Energy Frontier Research Center funded by the U.S. Department of Energy, Office of Science and Office of Basic Energy Sciences (DE-SC0001088).

# References

1. R. E. Merrifield, Vibronic states of dimers. *Radiation Res.* **20**, 154-158, (1963).

2. T. Förster, *Delocalized Excitation and Excitation Transfer* (Florida State University, Tallahassee, Florida, 1965), Vol. Bulletin No. 18, Division of Biology and Medicine, U.S. Atomic Energy Commission,

3. E. Charney, *The Molecular Basis of Optical Activity: Optical Rotatory Dispersion and Circular Dichroism* (John Wiley & Sons, New York, 1979),

4. W. Domcke, H. Köppel, and L. S. Cederbaum, Spectroscopic effects of conical intersections of molecular potential energy surfaces. *Molec. Phys.* **43**, 851-875, (1981).

5. A. Eisfeld, L. Braun, W. T. Stunz, J. S. Briggs, J. Beck, and V. Engel, Vibronic energies and spectra of molecular dimers. *J. Chem. Phys.* **122**, 134103-1-10, (2005).

6. A. Eisfeld, J. Seibt, and V. Engel, On the inversion of geometric parameters from absorption and circular dichroism spectroscopy of molecular dimers. *Chem. Phys. Lett.* **467**, 186-190, (2008).

7. F. C. Spano, The spectral signatures of Frenkel polarons in H- and J-aggregates. *Acc. Chem. Res.* **43**, 429-439, (2010).

8. K. A. Kistler, C. M. Pochas, H. Yamagata, S. Matsika, and F. C. Spano, Absorption, circular dichroism, and photoluminescence in perylene diimide bichromophores: Polarization-dependent H- and J-aggregate behavior. *J. Phys. Chem. B* **116**, 77-86, (2012).

9. S. Polyutov, O. Kühn, and T. Pullerits, Exciton-vibrational coupling in molecular aggregates: Electronic versus vibronic dimer. *Chem. Phys.* **394**, 21-28, (2012).

10. V. Tiwari, W. K. Peters, D. M. Jonas, Electronic resonance with anticorrelated pigment vibrations drives photosynthetic energy transfer outside the adiabatic framework. *Proc Natl Acad Sci U S A* **110**, 1203-1208, (2013).

11. A. J. Halpin, P. J. M. , R. Tempelaar, R. S. Murphy, J. Knoester, T. L. C. Jansen, and R. J. D. Miller, Two-dimensional spectroscopy of a molecular dimer unviels the effects of vibronic coupling on exciton coherences. *Nature Chem.* **6**, 196-201, (2014).

12. H.-G. Duan, P. Nalbach, V. I. Prokhorenko, S. Mukamel, and M. Thorwart, On the origin of oscillations in two-dimensional spectra of excitonically-coupled molecular systems. *New Journal of Physics* **17**, 072002, (2015).

13. J. Lim, D. Paleček, F. Caycedo-Soler, C. N. Lincoln, J. Prior, H. von Berlepsch, S. R. Huelga, M. B. Plenio, D. Zigmantas, and J. Hauer, Vibronic origin of long-lived coherence in an artificial molecular light harvester. *Nature Comm.* **6**, 7755-1-7, (2015).

14. M. Kasha, H. R. Rawls, and M. A. El-Bayoumi, The exciton model in molecular spectroscopy. *Pure Appl. Chem.* **11**, 371-392, (1965).

15. R. L. Fulton, and M. Gouterman, Vibronic coupling. I. Mathematical treatment for two electronic states. *J. Chem. Phys.* **35**, 1059-1071, (1961).

16. R. L. Fulton, and M. Gouterman, Vibronic coupling. II. Spectra of dimers. *J. Chem. Phys.* **41**, 2280-2286, (1964).

17. L. Valkunas, D. Abramavicius, and T. Mančal, *Molecular Excitation Dynamics and Relaxation: Quantum Theory and Spectroscopy* (Wiley-VCH, 2013),

18. T. Azumi, and K. Matsuzaki, What does the term "vibronic coupling" mean? *Photochem. and Photobiol.* **25**, 315-326, (1977).

19. G. Fischer, *Vibronic Coupling: The Interaction Between the Electronic and Nuclear Motions* (Academic Press, London, 1984),

20. E. Collini, C. Y. Wong, C. Y. Wilk, P. M. G. Curmi, P. Brumer, and G. D. Scholes, Coherently wired light-harvesting in photosynthetic marine algae at ambient temerature. *Nature* **463**, 644-647, (2010).

21. A. Chenu, and G. D. Scholes, Coherence in energy transfer and photosynthesis. *Ann. Rev. Phys. Chem.* **66**, 69-96, (2015).

22. S. Blau, D. I. G. Bennett, C. Kreisbeck, G. D. Scholes, and A. Aspuru-Guzik, arXiv:1704.05449v1, 2017).

23. G. S. Engel, T. R. Calhoun, E. L. Read, T.-K. Ahn, T. Mančal, Y.-C. Cheng, R. E. Blankenship, and G. R. Fleming, Evidence for wavelike energy transfer trhough quantum coherence in photosynthetic systems. *Nature* **446**, 782-786, (2007).

24. E. Romero, R. Augulis, V. I. Novoderezhkin, M. Ferretti, J. Thieme, D. Zigmantas, and R. van Grondelle, Quantum coherence in photosynthesys for efficient solar-energy conversion. *Nature phys.* **10**, 676-682, (2014).

25. H.-G. Duan, V. I. Prokhorenko, R. J. Cogdell, K. Ashraf, A. L. Stevens, M. Thorwart, and R. J. D. Miller, Nature does not rely on long-lived electronic quantum coherence for photosynthetic energy transfer. *Proc Natl Acad Sci U S A* **114**, 8493-8498, (2017).

26. N. Killoran, S. F. Huelga, and M. B. Plenio, Enhancing light-harvesting power with coherent vibrational interactions: A quantum heat engine picture. *J. Chem. Phys.* **143**, 155102-1-10, (2015).

27. M. Mohseni, P. Rebentrost, S. Lloyd, and A. Aspuru-Guzik, Environment-assisted quantum walks in photosynthetic energy transfer. *J. Chem. Phys.* **129**, 174106-1-9, (2008).

28. M. Levitus, and S. Ranjit, Cyanine dyes in biophysical research: The photophysics of polymethine fluorescent dyes in biomolecular environments. *Quat. Rev. Biophys.* **44**, 123-151, (2011).

29. A. Mishra, R. K. Behera, P. K. Behera, B. K. Mishra, and G. B. Behera, Cyanines during the 1990s: A review. *Chem. Rev.* **100**, 1973-2011, (2000).

30. F. Würthner, T. E. Kaiser, and C. R. Saha-Möller, 75 years of J-aggregates. *Angew. Chem. Int. Ed.* **50**, 3376-3410, (2000).

31. U. Resch-Genger, M. Grabolle, S. Cavaliere-Jaricot, R. Nitschke, and T. Nann, Quantum dots versus organic dyes as fluorescent labels. *Nature Meths.* **5**, 763-775, (2008).

32. M. E. Sanborn, B. K. Connolly, K. Gurunathan, and M. Levitus, Fluorescence properties and photophysics of the sufoindocyanine Cy3 linked covalently to DNA. *J. Phys. Chem. B* **111**, 11064-11074, (2007).

33. W. Lee, D. Jose, C. Phelps, A. H. Marcus, and P. H. von Hippel, A Single-Molecule View of the Assembly Pathway, Subunit Stoichiometry and Unwinding Activity of the Bacteriophage T4 Primosome (Helicase-Primase) Complex. *Biochemistry* **52**, 3157 – 3170, (2013).

34. W. Lee, P. H. von Hippel, and A. H. Marcus, Internally Labeled Cy3 / Cy5 DNA Constructs Show Greatly Enhanced Photostability in Single-Molecule FRET Experiments. *Nucleic Acids Res* **42**, 5967 – 5977, (2014).

35. C. Phelps, W. Lee, D. Jose, P. H. von Hippel, and A. H. Marcus, Single-Molecule FRET and Linear Dichroism Studies of DNA 'Breathing' and Helicase Binding at Replication Fork Junctions. *Proc Natl Acad Sci U S A* **110**, 17320 – 17325, (2013).

36. P. H. von Hippel, N. P. Johnson, and A. H. Marcus, 50 Years of DNA 'Breathing': Reflections on Old and New Approaches. *Biopolymers* **99**, 923-954, (2013).

37. T. Holstein, Studies of polaron motion: Part I. The molecular-crystal model. *Ann. Phys.* **8**, 325-342, (1959).

38. J. Roden, A. Eisfeld, M. Dvořák, O. Bünermann, and F. Stienkemeier, Vibronic line shapes of PTCA oligomers in helium nanodroplets. *J. Chem. Phys.* **134**, 054907-1-12, (2011).

39. C. R. Cantor, P. R. Schimmel, *Biophysical Chemistry Part II: Techniques for the study of biological structure and function* (Freeman, New York, 1980), Vol. 2, Biophysical Chemistry,

40. N. Berova, and K. Nakanishi, in *Circular Dichroism: Principles and Applications*, edited by N. Berova, K. Nakanishi, and R. W. Woody (John Wiley & Sons, Inc., New York, 2000), pp. 337-382.

41. P. F. Tekavec, G. A. Lott, and A. H. Marcus, Fluorescence-detected two-dimensional electronic coherence spectroscopy by acousto-optic phase modulation. *J. Chem. Phys.* **127**, 214307, (2007).

42. A. Perdomo, J. R. Widom, G. A. Lott, A. Aspuru-Guzik, A. H. Marcus, Conformation and electronic population transfer in membrane supported self-assembled porphyrin dimers by two-dimensional fluorescence spectroscopy. *J. Phys. Chem. B* **116**, 10757-10770, (2012).

43. J. R. Widom, N. P. Johnson, P. H. von Hippel, A. H. Marcus, Solution Conformation of 2-Aminopurine (2-AP) Dinucleotide by Ultraviolet 2D Fluorescence Spectroscopy (UV-2D FS). *New Journal of Physics* **15**, 025028, (2013).

44. K. J. Karki, J. R. Widom, J. Seibt, I. Moody, M. C. Lonergan, T. Pullerits, A. H. Marcus, Coherent Two-Dimensional Photocurrent Spectroscopy in a PbS Quantum Dot Photocell. *Nature Comm.* **5**, 5869-1-7, (2014).

45. S. Mukamel, *Principles of Nonlinear Optical Spectroscopy* (Oxford University Press, Oxford, 1995),

46. H. Dong, and G. R. Fleming, Inhomogeneous broadening induced long-lived integrated two-color coherence photon echo signal. *J. Phys. Chem. B* **118**, 8956-8961, (2014).

47. H. van Amerongen, L. Valkunas, and R. van Grondelle, *Photosynthetic Excitons* (World Scientific, Singapore, 2000),

48. F. Sánchez-Bajo, and F. L. Cumbrera, The use of the pseudo-Voigt function in the variance method of x-ray line broadening analysis. *J. Appl. Cryst.* **30**, 427-430, (1997).

49. J. R. Widom, A. Perdomo-Ortiz, W. Lee, D. Rappoport, T. F. Molinski, A. Aspuru-Guzik, A. H. Marcus, Temperature-Dependent Conformations of a Membrane Supported 'Zinc Porphyrin Tweezer' by 2D Fluorescence Spectroscopy. *J. Phys. Chem. A* **117**, 6171 – 6184, (2013).

50. C. Phelps, B. Israels, D. Jose, M. Marsh, P. H. von Hippel, and A. H. Marcus, Using microsecond single-molecule FRET to determine the assembly pathways of T4 ssDNA binding protein onto model DNA replicatoin forks. *Proc Natl Acad Sci U S A* **114**, E3612-E3621, (2017).

51. R. H. Byrd, J. Nocedal, and R. A. Waltz, *KNITRO: An Integrated Package for Nonlinear Optimization.* (Springer-Verlag, Berlin, Germany, 2006), Large-Scale Nonlinear Optimization, 35-59.

52. J. Spiriti, J. K. Binder, M. Levitus, and A. van der Vaart, Cy3-DNA stacking interactions strongly depend on identity of the terminal basepair. *Biophys. J.* **100**, 1049-1057, (2011).

53. P. F. Aramendía, R. M. Negri, and E. San Román, Temperature dependence of fluorescence and photoisomerization in symmetric carbocyanines. Influence of medium viscosity and molecular structure. *J. Phys. Chem.* **98**, 3165-3173, (1994).

54. D. H. Waldeck, and G. R. Fleming, Influence of viscosity and temperature on rotational reorientation. Anisotropic absorption of 3,3'-diethyloxadicarbocyanine Iodide. *J. Phys. Chem.* **85**, 2614-2617, (1981).

55. S. Völker, *Relaxation Processes in Molecular Excited States* (Kluwer, Dordrecht, 1989),

56. M. Aydin, Ö. Dede, and D. L. Akins, Density functional theory and Raman spectroscopy applied to structure and vibrational mode analysis of 1,1',3,3'-tetraethyl-5,5',6,6'-tetrachloro-benzimidazolocarbocyanine iodide and its aggregate. *J. Chem. Phys.* **134**, 064325-1-12, (2011).

57. L. I. Markova, V. L. Malinovskii, L. D. Patsenker, and R. Häner, J- vs. H-type assembly: pentamethine cyanine (Cy5) as a near-IR chroptical reporter. *Chem. Comm.* **49**, 5298-5300, (2013).

58. F. Nicoli, M. K. Roos, E. A. Hemmig, M. Di Antonio, R. Vivie-Riedle, and T. Liedl, Proximity-induced H-aggregation of cyanine dyes on DNA-duplexes. *J. Phys. Chem. A* **120**, 9941-9947, (2016).

**Supporting Information for: Temperature-dependent conformations of exciton-coupled Cy3 dimers in double-stranded DNA**

Loni Kringle,[1] Nicolas P. D. Sawaya,[2] Julia Widom,[3] Carson Adams,[1] Michael G. Raymer,[4] Alán Aspuru-Guzik,[2] and Andrew H. Marcus[1,*]

[1.] Department of Chemistry and Biochemistry, Center for Optical, Molecular and Quantum Science, University of Oregon, Eugene, OR 97403, USA

[2.] Department of Chemistry and Chemical Biology, Harvard University, Cambridge, MA 02138, USA

[3.] Department of Chemistry, University of Michigan at Ann Arbor, Ann Arbor, Michigan 48109, United States

[4.] Department of Physics, Center for Optical, Molecular and Quantum Science, University of Oregon, Eugene, OR 97403, USA

[*]e-mail: ahmarcus@uoregon.edu

**Table SI.** Optimized values of the Hamiltonian parameters of the Cy3 monomer DNA construct at various temperatures, obtained from model fits of Eqs. (17), (2) and (3) to the absorption spectra. These calculations used the electric transition dipole moment (EDTM) $|\mu_{eg}^0|$ = 12.8 D. The parameters listed are the electronic transition energy $\varepsilon_{eg}$, the vibrational mode frequency $\omega_0$, the Huang-Rhys parameter $\lambda^2$, and the standard deviation of the Gaussian disorder function $\sigma_{I,mon}$.

| $T$ (°C) | $\varepsilon_{eg}$ (cm$^{-1}$) | $\omega_0$ (cm$^{-1}$) | $\lambda^2$ | $\sigma_{I,mon}$ (cm$^{-1}$) |
|---|---|---|---|---|
| 15 | 18,285 +40/−39 | 1,116 +99/−103 | 0.54 +0.07/−0.06 | 333 +16/−15 |
| 25 | 18,277 +38/−37 | 1,109 +88/−90 | 0.56 +0.06/−0.06 | 347 +15/−14 |
| 35 | 18,266 +36/−35 | 1,119 +82/−84 | 0.56 +0.06/−0.06 | 353 +14/−13 |
| 45 | 18,262 +36/−35 | 1,113 +82/−82 | 0.56 +0.06/−0.05 | 380 −14/ +13 |
| 55 | 18,280 +39/−38 | 1,124 +93/−96 | 0.55 +0.06/−0.06 | 380 +15/−14 |
| 60 | 18,289 +42/−41 | 1,107 +95/−98 | 0.54 +0.07/−0.06 | 365 +17/−15 |
| 65 | 18,301 +45/−45 | 1,103 +103/−107 | 0.54 +0.07/−0.07 | 367 +18/−16 |
| 70 | 18,308 +44/−43 | 1,100 +102/−100 | 0.56 +0.07/−0.06 | 388 +17/−16 |
| 75 | 18,323 +49/−48 | 1,072 +120/−120 | 0.54 +0.08/−0.07 | 376 +19/−17 |
| 85 | 18,309 +42/−41 | 1,091 +94/−92 | 0.56 +0.06/−0.06 | 399 +16/−15 |

**Table SII.** Davydov splittings of the $0-0$ and $1-0$ vibronic bands. Experimental values are determined from the energy difference between the optimized upper energy symmetric (+) and lower energy anti-symmetric (−) states within the vibronic band. For the $1-0$ vibronic band, only single-particle states are considered. Theoretical values are determined by the expression $DS_{v_t-0} = 2Je^{-\lambda^2}\lambda^{2v_t}/v_t!$, where $v_t$ phonon occupancy.

| $T$ (°C) | $J$ (cm$^{-1}$) | $DS_{0-0}$ (cm$^{-1}$) | | $DS_{1-0}$ (cm$^{-1}$) | |
|---|---|---|---|---|---|
| | | Theory | Experiment | Theory | Experiment |
| 15 | 529 | 616 | 532 | 333 | 327 |
| 25 | 514 | 587 | 511 | 329 | 307 |
| 35 | 497 | 567 | 499 | 318 | 297 |
| 45 | 483 | 552 | 488 | 309 | 289 |
| 55 | 467 | 538 | 480 | 296 | 285 |
| 60 | 449 | 524 | 469 | 283 | 280 |
| 65 | 362 | 422 | 392 | 228 | 225 |

**Table SIII**: Temperature-dependent vibronic band intensity ratios $I_\pm^{(0-0)}/I_\pm^{(1-0)}$ for the symmetric (+) and anti-symmetric (−) exciton bands of the (Cy3)$_2$ dimer DNA construct. The temperature-independent vibronic band intensity ratio for the Cy3 monomer DNA construct is $I^{(0-0)}/I^{(1-0)} = 1.60$. The perturbation theory prediction is given by Eq. (16), which assumes that the system resides in the weak exciton-coupling regime (i.e. $|J| \ll \lambda^2\hbar\omega_0 = 602$ cm$^{-1}$).

| $T$ (°C) | $J$ (cm$^{-1}$) | Anti-symmetric exciton (−) | | Symmetric exciton (+) | |
|---|---|---|---|---|---|
| | | Theory | Experiment | Theory | Experiment |
| 15 | 529 | 4.31 | 2.60 | 0.84 | 0.62 |
| 25 | 514 | 4.02 | 2.47 | 0.83 | 0.63 |
| 35 | 497 | 3.88 | 2.43 | 0.82 | 0.66 |
| 45 | 483 | 3.81 | 2.37 | 0.86 | 0.68 |
| 55 | 467 | 3.77 | 2.36 | 0.91 | 0.72 |
| 60 | 449 | 3.80 | 2.35 | 0.94 | 0.74 |
| 65 | 362 | 3.29 | 2.22 | 1.07 | 0.88 |

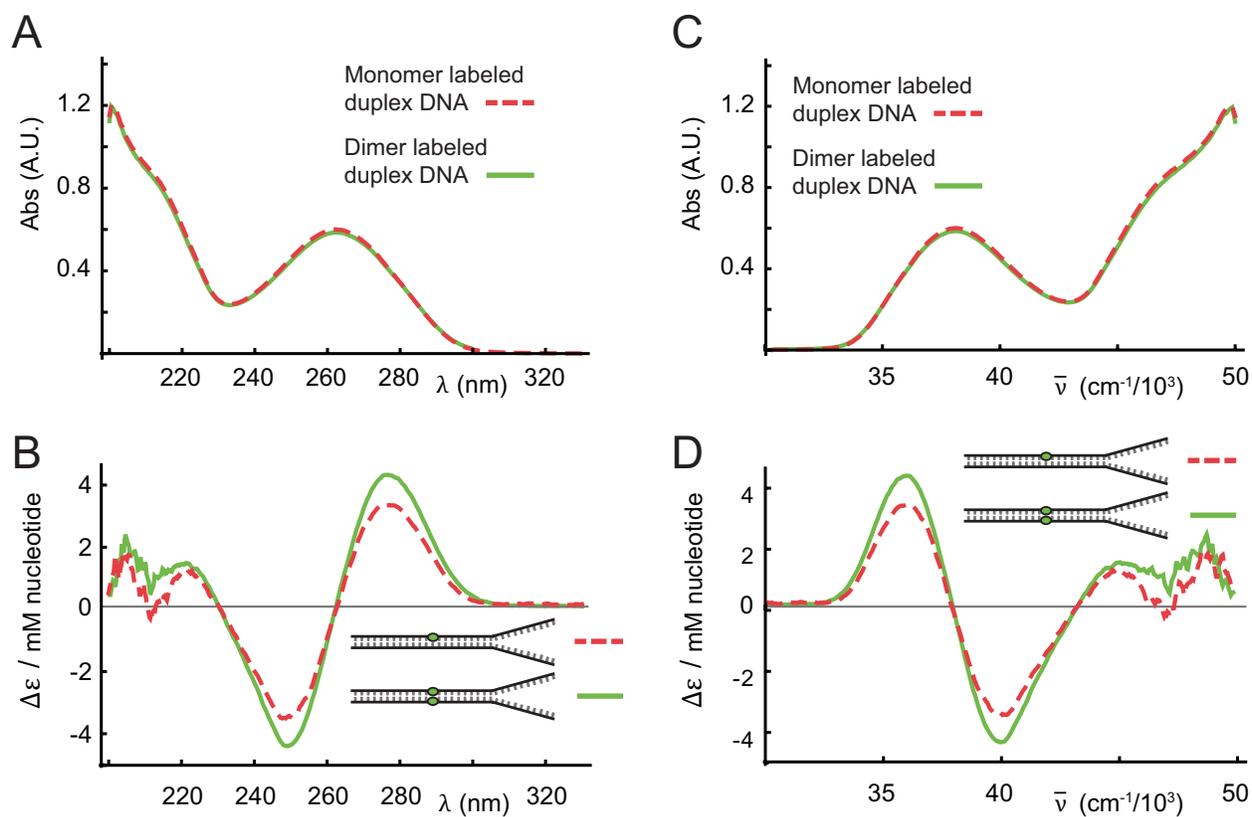

**Figure S1.** (*A* & *C*) Room temperature (25°C) absorption and (*B* & *D*) CD spectra for the Cy3 monomer DNA construct (dashed red), and the (Cy3)$_2$ dimer DNA construct (solid green), over the nucleobase spectral range. The spectra are shown as a function of the optical wavelength (*A* & *B*), and the wavenumber (*C* & *D*).

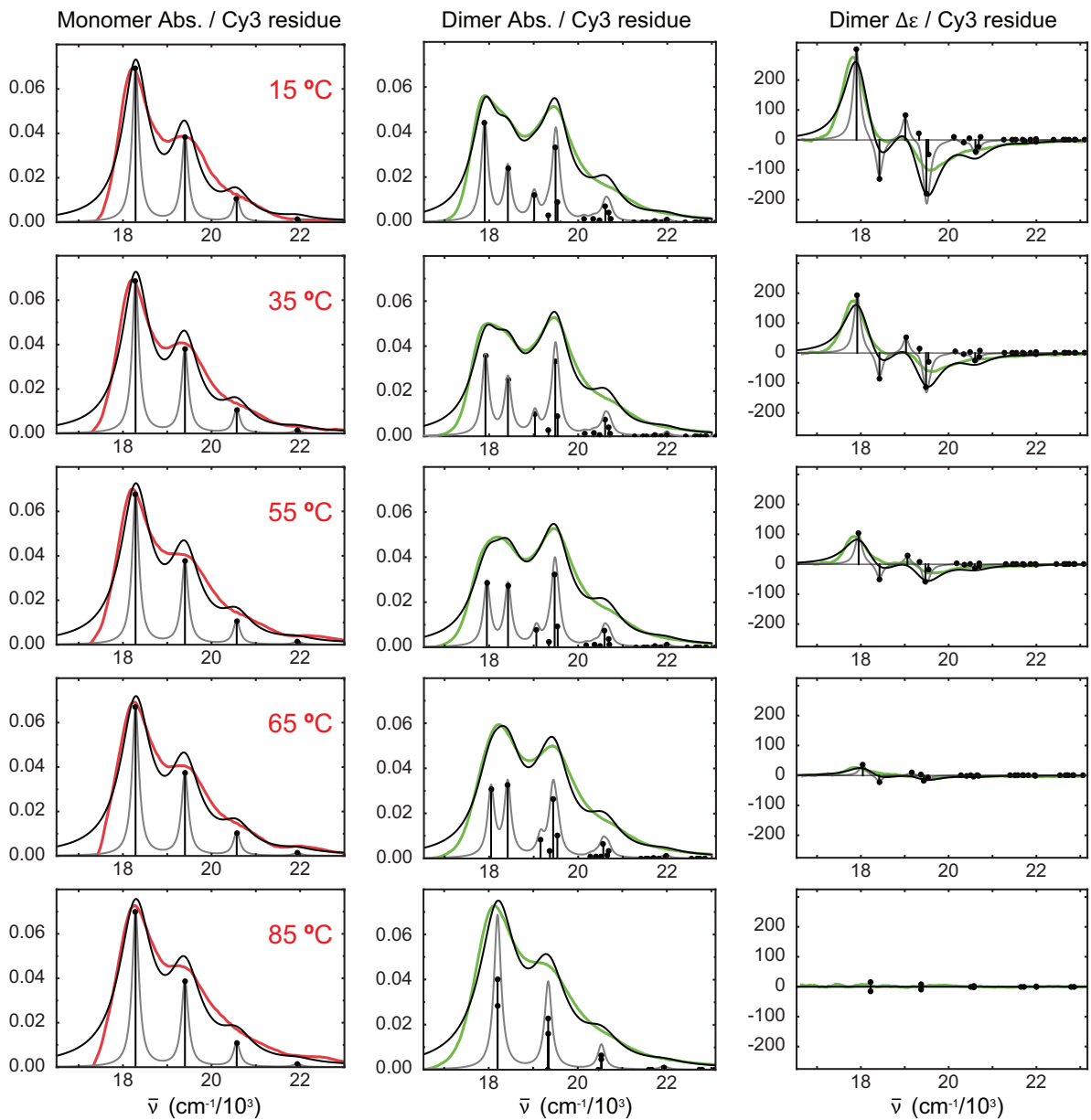

**Figure S2**. Temperature-dependent absorption spectra of the Cy3 monomer DNA construct (*left column, shown in red*), and the absorption (*middle column, in green*) and CD spectra (*right column, in green*) of the (Cy3)$_2$ dimer DNA construct (shown in green). Temperature increases from top to bottom, as indicated. Inhomogeneous lineshapes (black curves) are convolutions of the homogeneous Lorentzian lineshapes (grey curves) with the Gaussian static disorder functions [described by Eq. (3)]. The values for the temperature-dependent optimized parameters are listed in Table SI for the Cy3 monomer DNA construct, and in Table II for the (Cy3)$_2$ dimer DNA construct over the range 15 – 65°C.

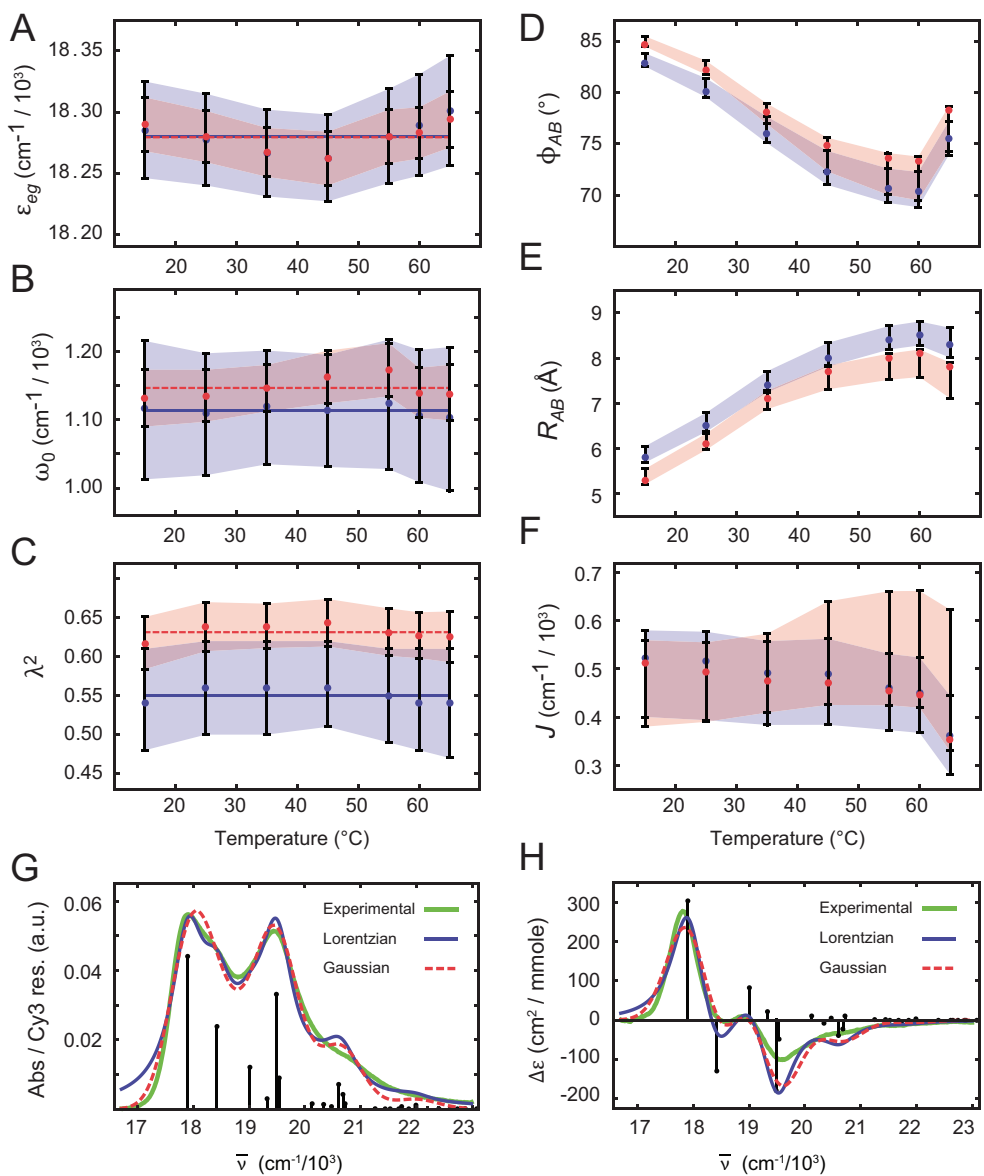

**Figure S3.** Comparison between optimized parameters based on the Lorentzian (blue) and Gaussian (red) models for the homogeneous lineshape. The intra-molecular parameters are (**A**) the electronic transition energy $\varepsilon_{eg}$, (**B**) the vibrational mode frequency $\omega_0$, and (**C**) the Huang-Rhys electronic-vibrational coupling parameter $\lambda^2$. The inter-molecular conformational parameters are (**D**) the inter-chromophore twist angle $\phi_{AB}$, (**E**) the inter-chromophore separation $R_{AB}$, and (**F**) the resonant coupling strength $J$. Error bars correspond to a 1% deviation of the target function from its minimum value. Comparison between experimental and simulated lineshapes for the absorption (**G**) and CD (**H**) spectra of the (Cy3)$_2$ dimer DNA construct at 15 °C.

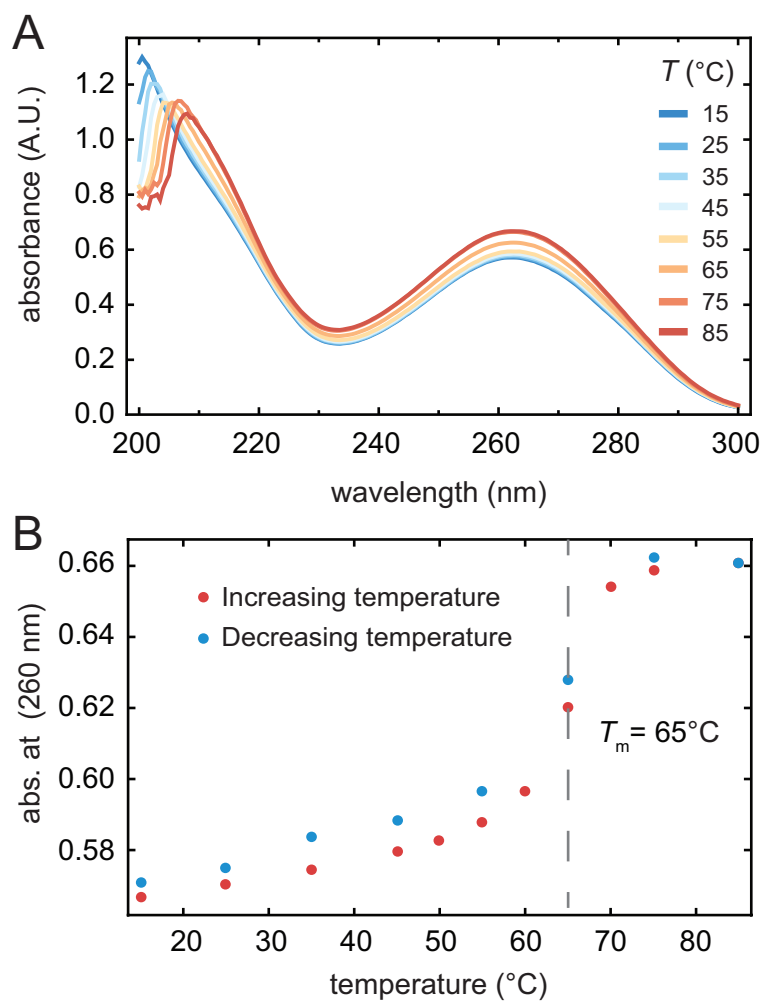

**Figure S4.** (*A*) Temperature-dependent absorption spectra of (Cy3)$_2$ dimer labeled dsDNA construct over the spectral range of the nucleobases. (*B*) Absorption intensity of at 260 nm as a function of temperature, showing the cooperative melting transition at $T_m$ = 65 °C.

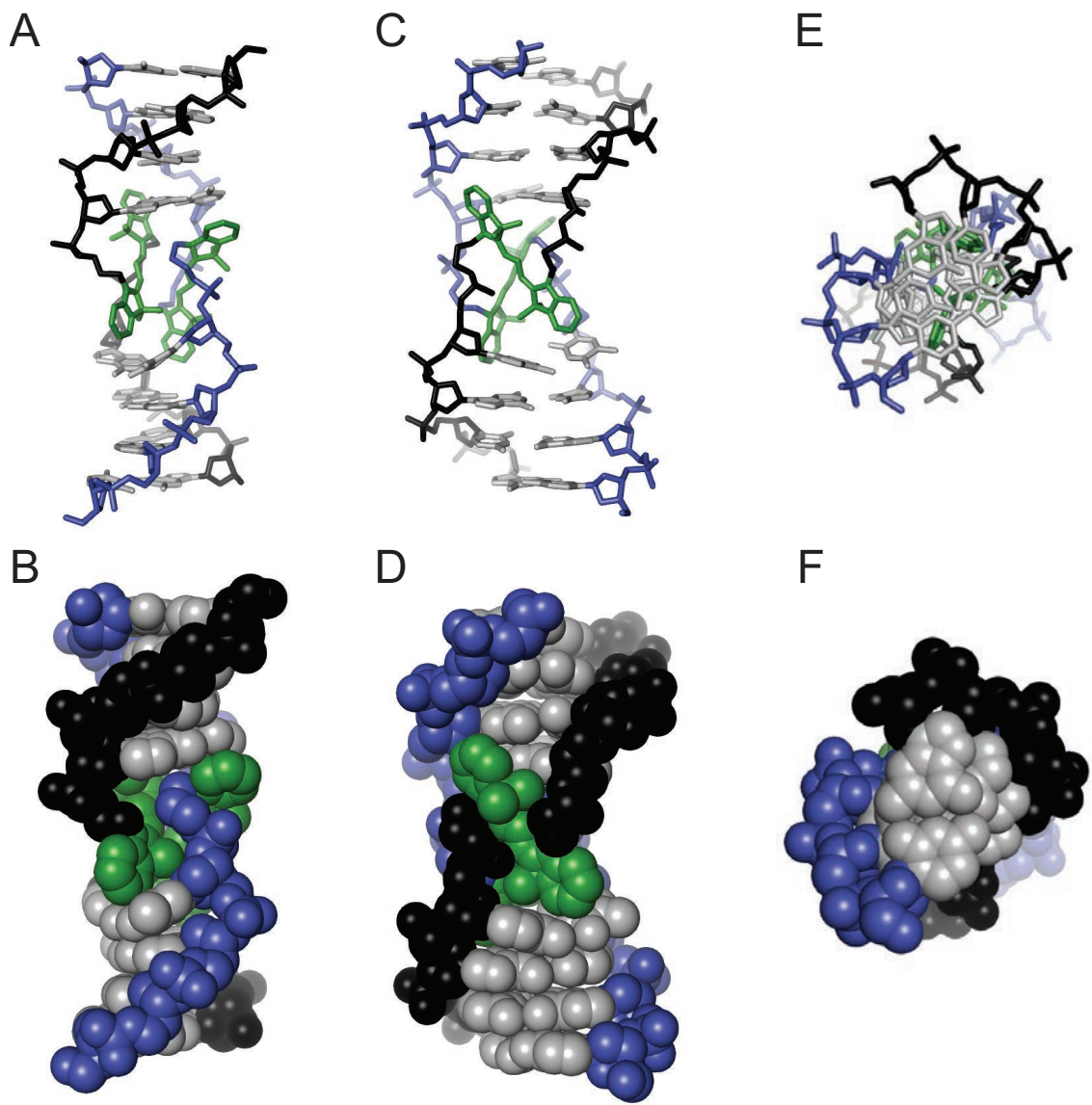

**Figure S5.** Ball and stick and space-filling models of the (Cy3)$_2$ DNA construct calculated using the Spartan program (Wavefunction, Inc.), and further refined using the PyMol program. The models show the (Cy3)$_2$ chromophore dimer, with approximate $D_2$ symmetry, within the surrounding bases of the DNA duplex. (*A* & *B*) The construct is shown with an orientation emphasizing the inter-chromophore separation, $R_{AB}$. (*C* & *D*) The construct is shown with an orientation an emphasizing the inter-chromophore twist angle $\phi_{AB}$. (*E* & *F*) The construct is viewed along the axial direction of the DNA duplex.